\documentclass{aa}

\usepackage{graphicx}
\usepackage{natbib}
\bibpunct{(}{)}{;}{a}{}{,} 
\usepackage{tikz}
\usepackage[toc,page]{appendix}
\usetikzlibrary{calc,tikzmark} 
\usepackage{hyperref}
\hypersetup{colorlinks=True,
    linkcolor=black,
    filecolor=black,      
    urlcolor=blue,
    citecolor=blue,
    }
    
\begin{document} 

   \title{Introducing the SHell misAlignment Detection for straylight Estimation (\textsc{SHADE}) algorithm: The case of \textit{XMM-Newton}}

   \author{S. Piscitelli$^{1,2}$ \and G. Ponti$^{2,3,1}$ \and M. Civitani$^2$  \and D. Spiga$^2$}

   \institute{$^1$ Como Lake Center for Astrophysics (CLAP), DiSAT, Università degli Studi dell’Insubria, via Valleggio 11, I-22100 Como, Italy \\ $^2$INAF – Osservatorio Astronomico di Brera, Via E. Bianchi 46, 23807 Merate, Italy \\ $^3$Max-Planck-Institut für extraterrestrische Physik, Giessenbachstraße 1, 85711 Garching, Germany}

   \date{Received 28 April 2025 / Accepted 24 September 2025 }

\abstract{When performing X-ray observations with a Wolter-I telescope, the presence of bright off-axis sources can introduce unfocused rays, known as stray light, which contaminate the detector and compromise the scientific analysis. Among the different components of stray light, single reflections off the hyperboloid section of the mirror shells often produce arc-like patterns on the detector. These arcs depend on not only the off-axis angle of the source but also the geometrical alignment of the individual shells. In this paper, we introduce the {SHell misAlignment Detection for stray light Estimation} (SHADE) algorithm, a novel and flexible tool designed to infer the misalignment parameters of individual shells, reproduce the geometry of stray light arcs, and predict its pattern on the detector. SHADE allows us to model each shell displacement with two parameters, $(\gamma,\xi)$, which represent the tilt amplitude and direction. While the algorithm is general and applicable to any Wolter-like telescope, we demonstrate its effectiveness using a set of \textit{XMM-Newton} observations of the low-mass X-ray binary GX5-1. As a proof of concept, we recover the best-fit misalignment parameters for a selected shell, obtaining $\gamma = 21.9''^{+10.3}_{-9.02}$ and $\xi = 5.88^{+1.02}_{-0.97}$ rad. SHADE represents a new approach to diagnosing mirror misalignments from stray light patterns and can support both pre- and postlaunch calibration efforts and future telescope designs.}

   \keywords{straylights, low mass x-ray binary, ray tracing, XMM-Newton}

\titlerunning{Introducing SHADE}
\authorrunning{S. Piscitelli, G. Ponti, M. Civitani, D. Spiga}
   \maketitle  

\section{Introduction}

X-ray telescopes are fundamental tools for investigating high-energy astrophysical processes. Their optical modules are typically based on double reflection systems, such as the Wolter-I design \citep{wolter1952spiegelsysteme}, where each mirror is composed of a parabolic and a hyperbolic surface. When a bright source is present just outside the field of view, rays may undergo single instead of double reflection \citep{de1999x}, producing stray light that contaminates the image recorded by the detector. This effect is particularly pronounced for rays reflected by the hyperboloid, as they often form characteristic arc-like patterns on the detector \citep{de1999x,spiga2015analytical}, compared to those reflected on the parabola, which typically do not reach the detector area.

\begin{figure}
    \centering
    \includegraphics[scale=0.6]{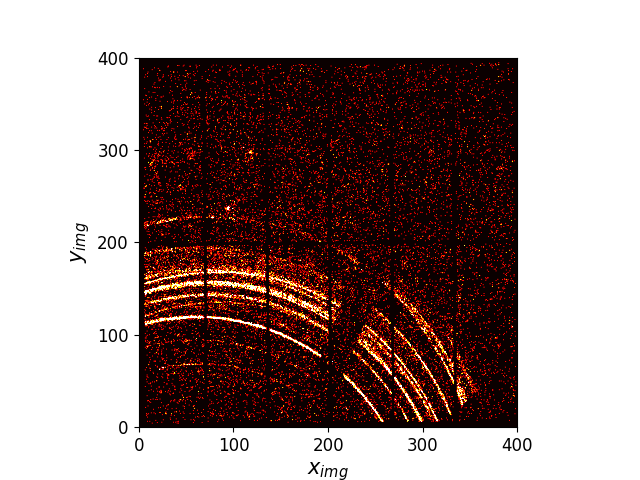}
    \caption{Observation 0932201101 performed around the LMXB GX5-1, showing an example of the presence of the stray light arcs for X-ray observations performed by Wolter I telescope with an off-axis source. The axes refer to pixel coordinates.}
    \label{fig:0932201101}
\end{figure}

An example of this phenomenon is shown in Fig. \ref{fig:0932201101}, which depicts an observation near the Low-Mass X-ray Binary \textcolor{black}{(LMXB)} GX5-1 with \textit{XMM-Newton}. In this work, we refer to $x_{img}$ and $y_{img}$ as the pixel coordinates. The angular distance $\theta$ between the source and the telescope's optical axis is a key parameter, since stray light contamination typically arises for sources located at off-axis angles between 20' and 78' \citep{de1999x}. In addition to the source position, the geometric disposition of the mirror shells plays a crucial role in shaping the stray light pattern, as shown in Appendix \ref{appendix:AppendixB}.

Previous studies have primarily focused on quantifying the stray light effective area under various conditions \citep{spiga2015analytical}, based on a limited number of vignetting parameters characteristic of the mirror module design \citep{spiga2011optics}. These models are built under the assumption of mirrors with no deformations and with optical axes aligned with each other. Observations frequently reveal deviations from the ideal pattern \citep{ponti2015,freyberg2006} -- which remain a matter of discussion -- resulting in the absence of a method that predicts the shape of the arcs. 

\begin{figure}
    \centering
    \includegraphics[scale=0.3]{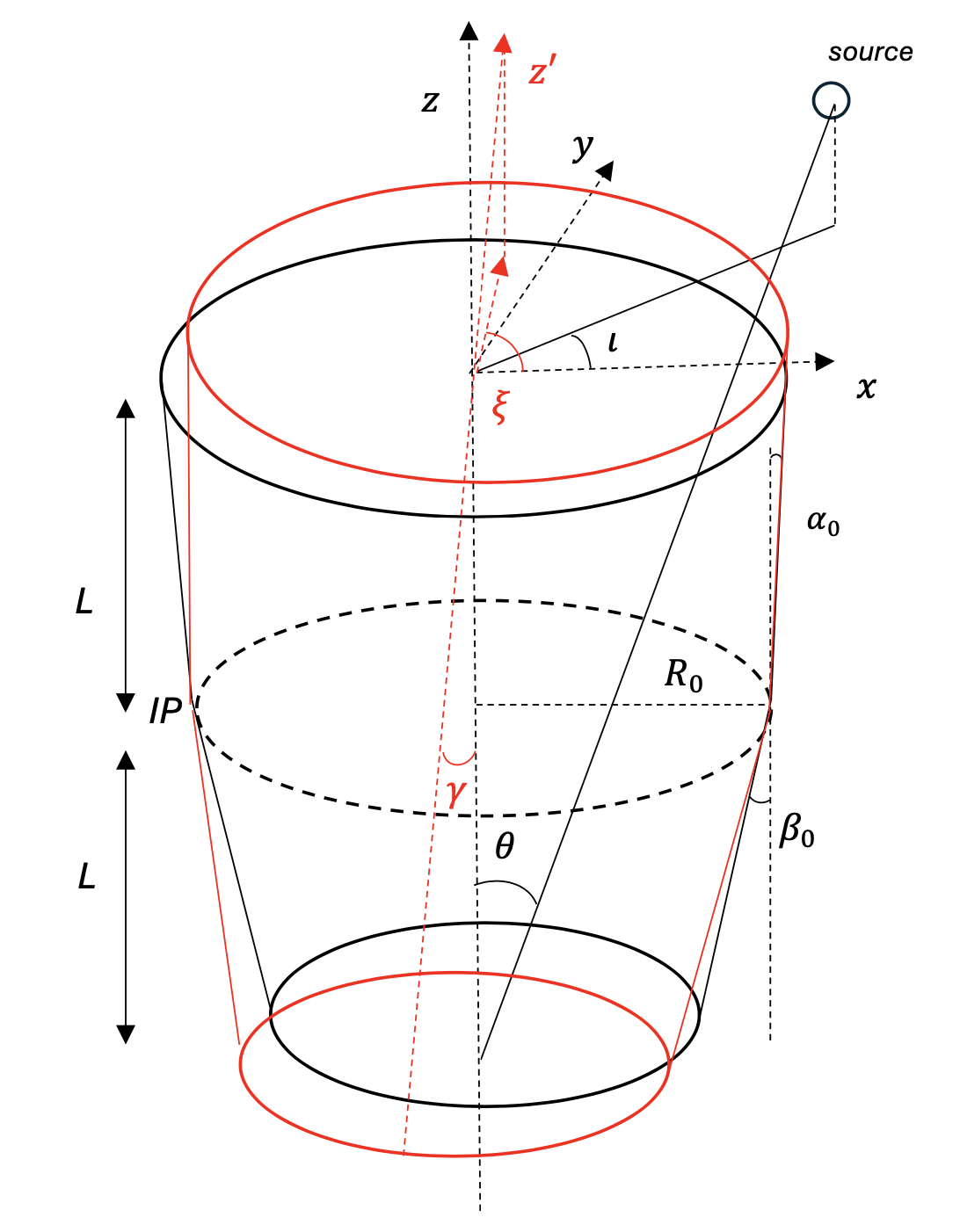}
    \caption{Representation of a sketch of a tilted optical axis. The tilted shell orientation is specified by the angles $\gamma$ and $\xi$, in cylindrical coordinates. The source direction is specified by the angles $\theta$, $\iota$ in the same reference frame.}
    \label{fig:tilt1}
\end{figure}

In this work, we present the SHell misAlignment Detection for straylight Estimation (SHADE) algorithm, designed to reproduce the observed arc geometry by inferring the misalignment and shift parameters of individual mirror shells from observed stray light arcs. The relevance of SHADE lies in its ability to infer the misalignment parameters of the mirrors by deriving them from stray light, playing an important role in pre- and post-calibration processes by predicting and reproducing the stray light pattern. Although simulations based on ray tracing can reproduce the pattern, they cannot easily perform parameter estimations of misaligned mirrors due to the uncertainty related to the simulated rays and the lack of a criterion used to compare simulations and the observed pattern. A possible path for comparison would be to interpolate the points from both arcs separately, including the inner variability of the positions of the rays through ray tracing. This effect is smoothed by the analytical approach, supporting its application. Also, to avoid poorly interpolated curves, the only way to infer these parameters with ray tracing would be to perform a substantial amount of simulations and choose the most significant arc among the ones generated. This task can be time-consuming; by contrast, an analytical approach does not present this obstacle. Once the tilt parameters are estimated, SHADE enables a more accurate modeling of the stray light contamination. Each shell misalignment is parameterized by the angle formed between the axis of the shell and the optical assembly axis, $\gamma$, and its orientation, $\xi$, on the aperture plane, as shown in Fig. \ref{fig:tilt1}. We applied this algorithm to a set of \textit{XMM-Newton} observations of the bright X-ray source GX5-1 \citep{bhulla2019astrosat}, selected for their suitable data quality, aimpoint stability, and position angle (PA) coverage.

This paper is organized as follows: In Sect. \ref{sect1}, we describe the geometrical properties of stray light patterns with a focus on \textit{XMM-Newton} observations. We introduce a baseline routine to simulate the arcs using analytical methods. In Sect. \ref{sect2}, we present the modeling of shell misalignments and their implementation within the simulation routine. In Sect. \ref{sect3}, we outline the SHADE pipeline and apply the algorithm to a selected sample of \textit{XMM-Newton} observations.

\section{The stray light pattern in \textit{XMM-Newton}\label{sect1}}

\subsection{Observations considered in this work}

\begin{figure}
    \centering
    \includegraphics[scale=0.6]{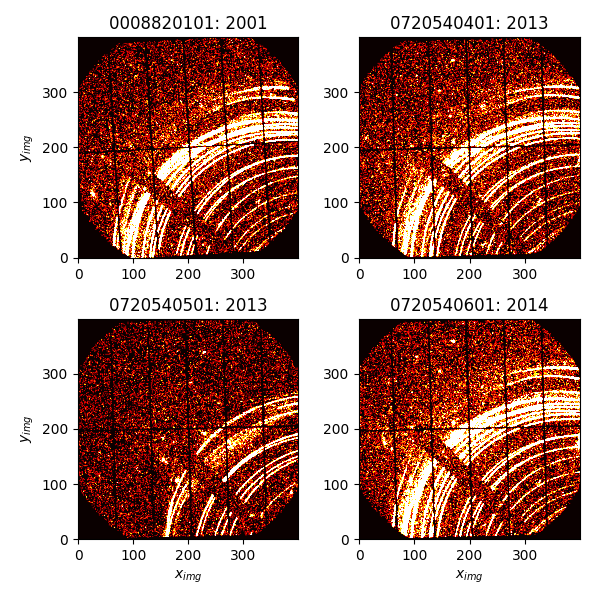}
    \caption{Observations considered in the analysis of the stray light arcs. The quality of these observations depends on the shared aimpoint and relationship between the PA values. Also, the different time span covered helps to explore the time variability of the phenomenon.}
    \label{fig:obsgrid}
\end{figure}

The observations considered in this work are shown in Fig. \ref{fig:obsgrid} and summarized in Table \ref{tab:infobs}. This sample is particularly suitable for our analysis for two main reasons. First, the observations share the same aimpoint despite spanning a time interval of 13 years, offering an ideal setup to investigate the consistency of the arc geometry over time. Second, the sample exhibits a wide range of PAs, including the observation 0720540501, which is characterized by an orientation nearly opposite to the others. This diversity allows us to study the stray light pattern under different observational configurations. A discussion on the implications of the PA in \textit{XMM-Newton} observations, depending on the telescope's reference systems, is provided in Appendix \ref{appendix:AppendixA}.

\textit{XMM-Newton} is equipped with three different X-ray cameras for observations \citep{schartel2024xmm}. In this work, we focus on data acquired with the EPIC-PN camera. This choice is motivated by the larger effective area of EPIC-PN compared to the EPIC-MOS cameras \citep{struder2001european}. Furthermore, the EPIC-MOS cameras (M1 and M2) have suffered damage in recent years \citep{damage}, leading to the loss of functionality in some Couple Charged Devices (CCDs). This degradation further supports the preference for EPIC-PN data, as it allows for a cleaner visualization of the stray light arcs.

\begin{figure}
    \centering
    \includegraphics[scale=0.2]{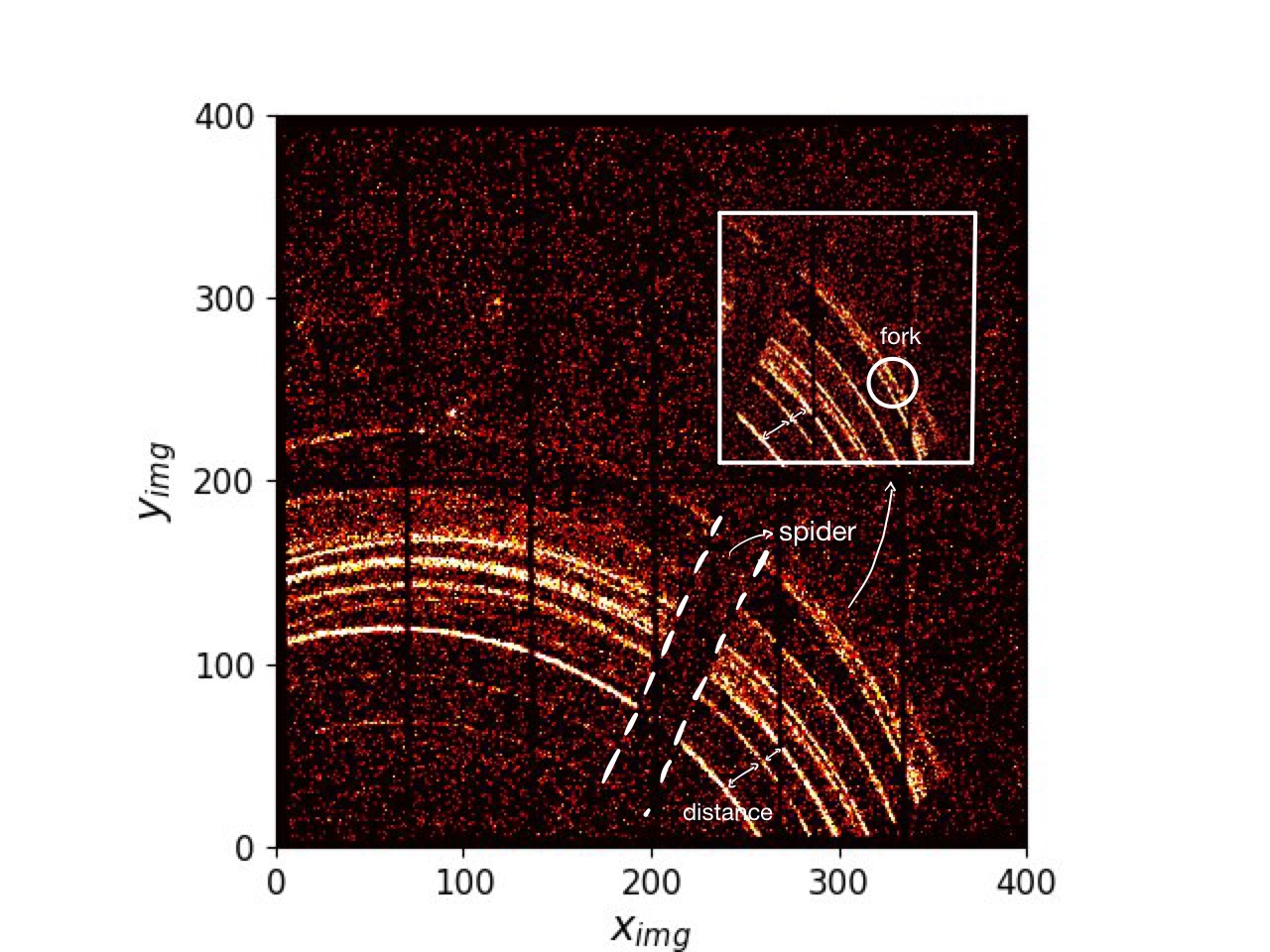}
    \caption{Visual representation of the geometrical features of the arcs reproduced in this study, including bifurcations and different spacing among shells.}
    \label{fig:0932201101_sketch}
\end{figure}

In addition to this sample, we refer to observation 0932201101 (Fig. \ref{fig:0932201101}) as a representative case to illustrate the geometrical features of the stray light arcs we aimed to reproduce. This observation was also used as a testbed to validate our modeling approach. A close-up inspection of the stray light pattern in Fig. \ref{fig:0932201101_sketch} reveals several distinctive characteristics. First, the arcs exhibit bifurcations (or “forks”) where portions of the stray light pattern intersect. Notably, this behavior is not observed for all shells. Second, the radial separation between adjacent arcs is not uniform but varies along the pattern. These irregularities do not happen in a perfectly co-aligned mirror, which we refer to as the perfect optical module, suggesting that the observed pattern may result either from specific geometrical properties of the telescope or from individual differences among the shells that we aim to describe with our code.

\begin{table}
\caption{Coordinates and time of the main \textit{XMM-Newton} observations.}
\label{tab:infobs}
\centering
\scalebox{0.88}{
\begin{tabular}{|c|c|c|c|c|}
\hline
     & RA (deg) & DEC (deg) & PA (deg) & DATE \\
     \hline
   0008820101  & 270.966 & -24.3602 & 90.9 & 2001-03-08 \\
     \hline
   0720540401  & 270.968 & -24.3607 & 89.19 & 2013-03-08 \\ 
     \hline
   0720540501  & 270.968 & -24.3607 & 267.6 & 2013-09-03 \\ 
     \hline
   0720540601  & 270.968 & -24.3607 & 87.6 & 2014-03-05 \\ 
   \hline
\end{tabular}}
\newline
\end{table}

\subsection{Analytical simulation of stray light}
Previous studies \citep{spiga2015analytical, spiga2016x} reproduced the stray light arcs due to single reflections on the hyperbola of a single shell using the following parametric equations, which assume that the source is off-axis by an angle $\theta$ in the direction of the x-axis:
\begin{gather}
    x(\phi) = (R_0 - 2\beta_0 f)\cos\phi + \theta f \cos(2\phi) \label{params_eq_1}, \\
    y(\phi) = (R_0 - 2\beta_0 f)\sin\phi + \theta f \sin(2\phi). \label{params_eq_2}
\end{gather}
These equations describe the ray intersections with the focal plane, denoted by $(x,y)$, as a function of the polar angle $\phi \in [0, 2\pi]$, after a single reflection on the hyperbola of a given shell. A rotation by an angle, $\iota$, should be operated on the coordinates $(x, y)$ in order to align the off-axis source in the correct orientation, where $\iota$ is the polar angle of the source represented in Fig. \ref{fig:tilt1}. In this formulation, $f$ represents the focal length of the telescope, $R_0$ is the radius of the shell at the Intersection Plane (IP), defined as the plane $z=0$, with $z$ aligned along the optical axis of the telescope. Defining the parameter $\alpha_0$ as the slope of the parabola with respect to the optical axis, the slope of the hyperbola is given by $\beta_0 = 3\alpha_0$ \citep{spiga2015analytical,stramaccioni2000xmm}. This configuration is described in Fig. \ref{fig:tilt1}.

Properly reproducing the stray light pattern with an analytical method requires two clarifications related to the double cone approximation \citep{spiga2009analytical}. The first regards its interpretation, and the other the concavity of the mirrors.
The first clarification can be addressed by considering the cone passing through the two edges as a fairly representative model for single reflections on the hyperbolic segment alone, compared to the typical tangent to the hyperbola at the IP. This choice entails a greater angle via the following correction of the $\beta_0$ parameter:
\begin{equation}
    \beta_0^{CH} = \beta_0^{DC} + \delta\beta_1,
\end{equation}
where $\beta_0^{CH}$ refers to the chord angle, $\beta_0^{DC}$ is the usual double-cone slope, and $\delta\beta/\beta_0^{DC} \sim 1\%$ for the \textit{XMM-Newton} case. In this approximation, we refer to $\beta_0^{CH}$ as the $\beta_0$ used in Eqs. \eqref{params_eq_1} and \eqref{params_eq_2}.

\begin{figure}
    \centering
    \includegraphics[scale=0.55]{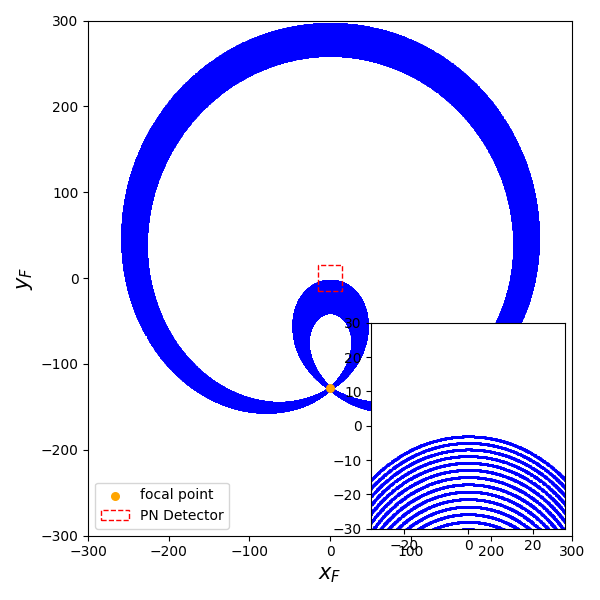}
    \caption{Stray light pattern generated with parametric Eqs \eqref{params_eq_1} and \eqref{params_eq_2} considering a source at $\theta = 1^\circ$ and $\iota = 3\pi/2$. The number of shells generated is limited due to the geometric configuration of the telescope.}
    \label{fig:analytic_pattern}
\end{figure}

The second consideration is that the double cone provides a focal length that exceeds the actual Wolter-I focal length for the same shell geometry. Due to the concavity of the optical surfaces, the incidence angles of the mirrors in a Wolter-I telescope vary along the $z$-axis, leading to a systematic overestimation of these angles on the hyperbolic segment, such that $\beta^{W}_{0}(z) \geq \beta^{DC}_{0}$ \citep{spiga2009analytical}, where $\beta^{W}_{0}(z)$ represents the Wolter-I incidence angles. 
The chord approximation introduces a different discrepancy compared to the double cone, resulting in 
$\beta_{0} \geq \beta^W_{0}(z)$. To account for this discrepancy, a correction factor is applied to the nominal grazing incidence angle $\beta_0$, with an additional adjustment of $\delta \beta_2 / \beta_0^{DC} \sim 1\%$
for \textit{XMM-Newton} \citep{spiga2009analytical} to calculate the focal length in the chord approximation $f$ from the focal length of the Wolter configuration $f_W$:
\begin{equation}
    f = f_{W} \frac{\tan(\beta^{DC}_0)}{\tan(\beta^{CH}_0 +\delta\beta_2)}.
\end{equation}
These corrections ensure consistency between the analytical model and ray-tracing simulations based on the Wolter I model, resulting in a focal length of $f=7350$ mm.
With the values taken for $R_0$ in the \textit{XMM-Newton} module, Fig. \ref{fig:analytic_pattern} shows the simulated stray light pattern using Eqs. \eqref{params_eq_1} and \eqref{params_eq_2} for a source at $\theta=1^\circ$ and $\iota = 3\pi/2$. In the simulation, we considered only the shells whose stray light intersected the detector area. The reasons for the limited number of shells visualized on the detector are discussed in more detail in Appendix \ref{appendix:AppendixB}.  
We highlight that the analytical simulation in Fig. \ref{fig:analytic_pattern} assumes that all mirror shells are perfectly aligned. Under this assumption, the simulated pattern does not reproduce the bifurcations, overlaps, and nonuniform radial spacing observed in Fig. \ref{fig:0932201101_sketch}. These discrepancies leave room for investigating the influence of misalignments in individual shells, motivating the development of a more flexible approach capable of accounting for these geometric deviations.

\section{Tilts of the shells\label{sect2}}

\subsection{Misalignments}
Assuming a tilt for the hyperbola implies that the incident angles between the light and the hyperbola for tilted and non-tilted mirrors differ. This geometrical feature can be described by two parameters: the amplitude of the angle between a light ray and the slope angle of the hyperbola, named $\gamma$, and a polar angle, $\xi$, which describes the direction of the tilt on the detector plane, defined with respect to the detector x-axis, as shown in Fig. \ref{fig:tilt1}. The orientation of the detector x-axis is set by the PA relative to the North Celestial Pole (\textcolor{black}{NCP}) in the equatorial coordinate system and is mentioned in Appendix \ref{appendix:AppendixA}. Since the detector reference frame rotates with the PA, we highlight that the angle, $\xi$, is defined with respect to the position of the detector x-axis when PA=$3\pi/2$.
In previous studies, \citep{stockman2001environmental,marioni1999xmm} the maximum misalignment of the shells registered is $30''$, due to the precision imposed by a vertical optical bench used to arrange the structure. However, misalignments of the hyperbola up to $1'$ have also been observed \citep{stockman1999xmm,glatzel1994assembly}.
 
In order to extend the simulation to misaligned shells, we derived the equations that represent the stray light pattern of a tilted shell, obtaining additional terms to add to Eqs. \eqref{params_eq_1} and \eqref{params_eq_2}:
\begin{gather}
    x'(\phi) = x(\phi) - \gamma f[\cos (2\phi - \xi + \iota) + \cos(\xi-\iota)]  \label{eqmis1}, \\
    y'(\phi) = y(\phi) - \gamma f[\sin (2\phi - \xi + \iota) + \sin(\xi-\iota)]. \label{eqmis2}
\end{gather}
The complete derivation of these equations is reported in Appendix \ref{appendix:AppendixC}. We note that these equations correctly reduce to Eqs. \eqref{params_eq_1} and \eqref{params_eq_2} when $\gamma = 0$. Moreover, the case of a shell misaligned in the off-axis plane corresponds to $\xi = 0$, whereas the case $\xi = \iota$ means that the shell misalignment lies exactly in the off-axis plane. As in the case of Eqs. \eqref{params_eq_1} and \eqref{params_eq_2}, a rotation by an angle $\iota$ should be operated on the coordinates $(x,y)$ in order to align the source with its actual direction.

\begin{figure}[h]
    \centering
    \includegraphics[scale=0.3]{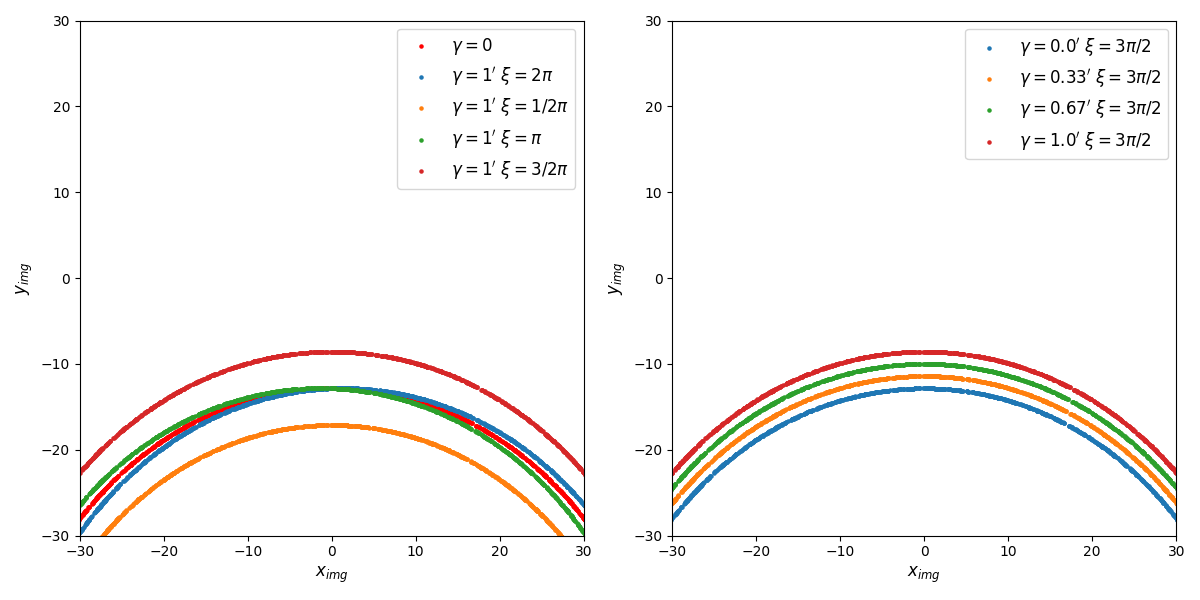}
    \caption{Illustration of a $1'$ misalignment in different directions $\xi$ (left panel), relative to a source located at $\iota = 3\pi/2$, and the effects of varying the amplitude (right panel) while keeping the direction $\xi$ fixed. We observe that the shell tilt can significantly affect the shape of the arc in the resulting image.}
    \label{fig:mis_grid}
\end{figure}

\begin{figure}
    \centering
    \includegraphics[scale = 0.55]{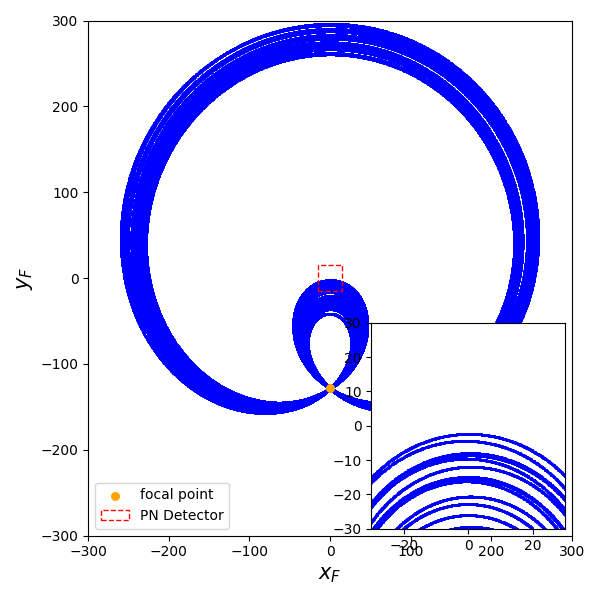}
    \caption{Simulation of the stray light arcs assuming uniformly sampled values for the parameters modeling the tilt in the intervals $\gamma \in [0,1]'$ and $\xi =\in [0,2\pi]$. This simulation yields a number of intersections between single-reflection traces, as expected from the \textit{XMM-Newton} observation.}
    \label{fig:simtilted}
\end{figure}

Analyzing the behavior of the pattern in response to different orientations of the tilt is a key aspect for extracting useful physical considerations. We performed the simulation of the single arc for shell 16 using Eqs. \eqref{eqmis1} and \eqref{eqmis2} with $\theta = 1^\circ$ and $\iota = 3\pi/2$. We considered the number of shells, counting from the one with the smallest $R_0$ (shell 58) to the largest (shell 1). As a first example, we considered four different amplitudes, equally spaced between 0 and $1'$, for the misalignment, keeping $\xi$ fixed at $3\pi/2$, and compared these with the zero tilt configuration. Secondly, we fixed the amplitude $\gamma = 1'$ and considered four different orientations of the misalignments between 0 and $2\pi$.
The result can be observed in Fig. \ref{fig:mis_grid}, where the presence of the tilt changes the position of the arc in different ways according to $\xi$ and in different magnitude according to $\gamma$.
A combination of these parameters can reproduce the features anticipated in Fig. \ref{fig:0932201101_sketch}, as shown in Fig. \ref{fig:simtilted} where we implemented in the previous simulation values for $\gamma$ and $\xi$ drawn from the uniform distributions in the intervals $\gamma = [0,1]'$ and $\xi = [0,2\pi]$.

\subsection{Shift of the shells}
\begin{figure}
    \centering
    \includegraphics[scale=0.3]{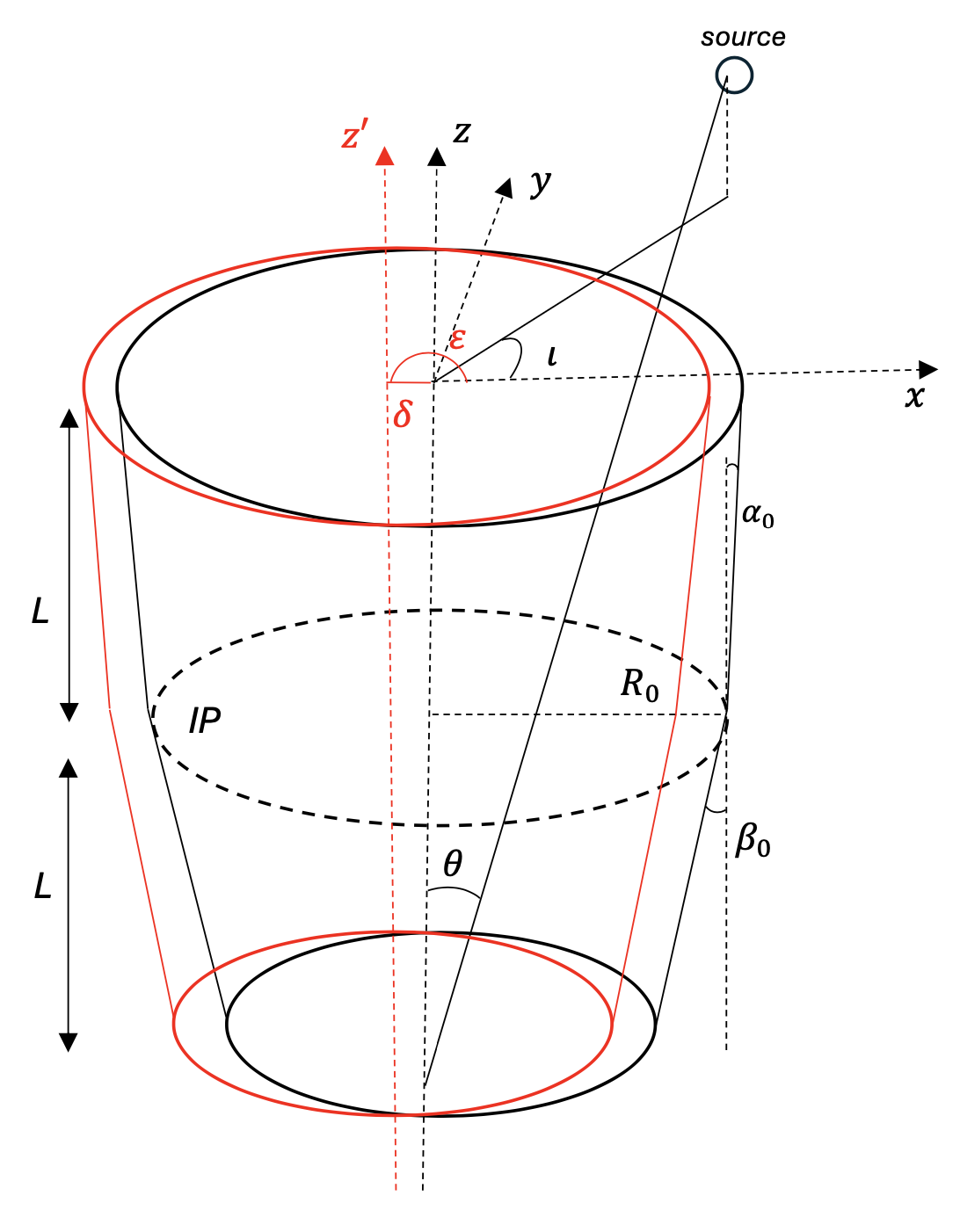}
    \caption{Sketch of the optical axis shift. The shift amplitude is labeled $\delta$, whereas the shift orientation on the aperture plane is labeled $\epsilon$.}
    \label{fig:tilt2}
\end{figure}

We now investigate the impact of a shift in the optical axis caused by a misplacement of the mirror module during the integration process. Figure \ref{fig:tilt2} illustrates an example of such a shift, which can be described by introducing two additional parameters, $(\delta,\epsilon)$, representing the distance between the theoretical and the actual optical axis as well as its direction in the aperture plane. Documentation related to the integration process \citep{dechambure1997producing} provides a threshold of $\delta < 5 \hspace{1mm} \mu\mathrm{m}$ as the maximum allowable shift.

A shift in the optical axis modifies the source coordinates as 
\begin{gather} 
\theta' = \sqrt{\theta^2 + \left(\frac{\delta}{f}\right)^2 - 2\theta\left(\frac{\delta}{f}\right)\cos(\epsilon-\iota)} \label{shift1}\ ,\\
\iota' = \arctan\left(\frac{\theta\sin\iota + \left(\frac{\delta}{f}\right)\sin\epsilon}{\theta\cos\iota + \left(\frac{\delta}{f}\right)\cos\epsilon}\right) \label{shift2}, 
\end{gather}
where $f$ indicates the focal length of the telescope. The derivation of these equations is tackled in Appendix \ref{appendix:AppendixC}.
Assuming the maximum value for $\delta = 5 \mu\mathrm{m}$, this contribution becomes negligible in the case of \textit{XMM-Newton}, where $\delta/f \sim 10^{-7}$, compared to $\theta$ expressed in radians.

\section{The Shell misAlignment Detection for straylight Estimation \label{sect3}}

\subsection{Time invariance of stray light patterns}
The reproducibility of stray light arcs in X-ray observations relies on the assumption of time invariance of the geometric configuration of the mirror module. This hypothesis is supported by the comparison between the different observations shown in Fig. \ref{fig:obsgrid}, which were recorded at similar orientations of the telescope. We performed a statistical comparison using a chi-squared test applied to various pairs of observations. Each image was divided into bins based on pixel intensity values, and histograms were created to represent the number of pixels in each intensity range. After normalizing the histograms to obtain probabilities, we compared the distributions of pixels from different images using the $\chi^2$ function from the SciPy.stats module \citep{2020SciPy-NMeth}. This method provides a quantitative assessment of the similarities between images. The test yields a p-value $p \approx 1$ \citep{wasserstein2016asa} for all pairs of images, supporting the assumption that stray light patterns remain stable over time. 

\begin{figure}
    \centering
    \includegraphics[scale=0.3]{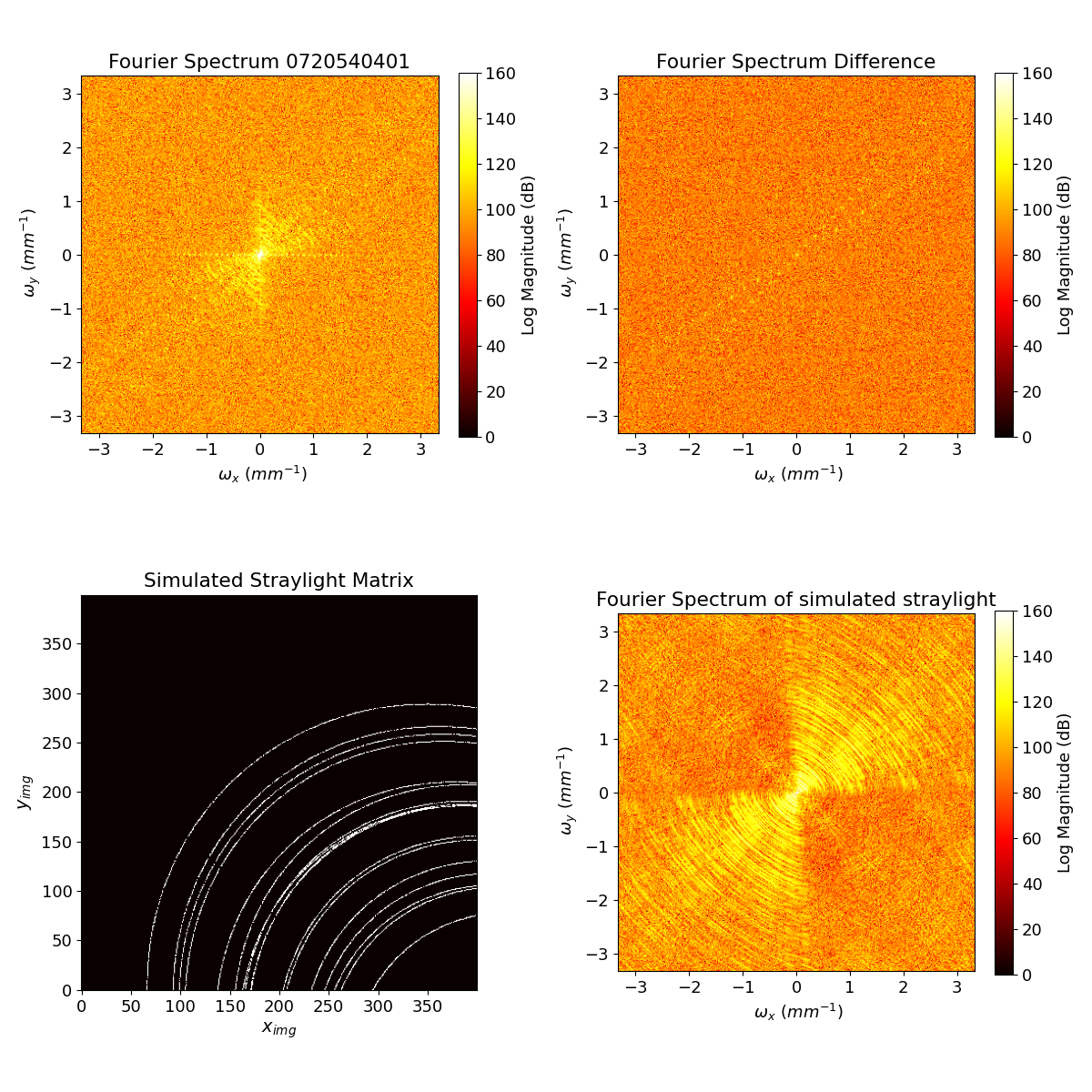}
    \caption{Fourier spectrum of the observation 0720540401 (top-left panel) and the spectrum of the matrix resulting from the difference between 0720540401 and 0720540601 after being properly rotated according to their PAs (top-right panel). The top-left panel shows particular features associated with the stray light with weak counterparts in the right panel. These arc-shaped features were reproduced by applying the Fourier transform to a simulated matrix with uniformly distributed tilt parameters (bottom-left panel), producing the plot shown in the bottom-right panel. $\omega_x$ and $\omega_y$ represent the frequencies along the two axes.}
    \label{fig:fourier}
\end{figure}

This method assesses global consistency between images, while locality can be tested both visually -- by subtracting the pixels of the images after a proper rotation according to their PAs, which results in an almost complete disappearance of the stray light pattern -- and in the Fourier space. Since the stray light in \textit{XMM-Newton} detectors is characterized by arcs covering almost half of the image, we expected this pattern to manifest in the Fourier spectrum with specific frequencies. In this context, frequencies -- indicated as $\omega_x$ and $\omega_y$ -- represent the scale of variability in luminosity of the pixels in the image. Applying the same Fourier transformation, using the scipy fast Fourier transform (FFT) \citep{virtanen2021scipy}, to the matrix of residuals provides almost no features associated with the stray light observed in the Fourier spectrum of the observations, supporting local invariance. A comparison between the top-left panel and the bottom-right panel in Fig. \ref{fig:fourier} shows that the features of the stray light pattern follow the same arc shape in the frequency domain, since the axes in this plot also represent the spatial direction of the frequencies in the observation. Figure \ref{fig:fourier} shows the frequencies of the spectrum, expressed in $mm^{-1}$, along the two axes, indicating the relevant frequencies for each line of pixels in both $(x,y)$ directions. The amplitude of the mode was considered in decibels. The spectrum was also shifted, placing the lower frequencies in the center for better visualization.

In the end, we used a convolutional neural network \citep{simonyan2015deepconvolutionalnetworkslargescale} to add another similarity criterion between observations. This criterion is based on the extraction of essential local features from images and assigns a similarity metric on a scale from 0 to 1, where 1 describes identical images. We applied this algorithm to all three pairs of observations -- 0008820101, 0720540401, and 0720540601 -- obtaining a similarity score of $\sim0.96$ for each of them.

Validating the time invariance of stray light patterns further supports the use of multiple observations to strengthen the association process.

\subsection{Associations of arcs}
\begin{figure}
    \centering
    \includegraphics[scale=0.25]{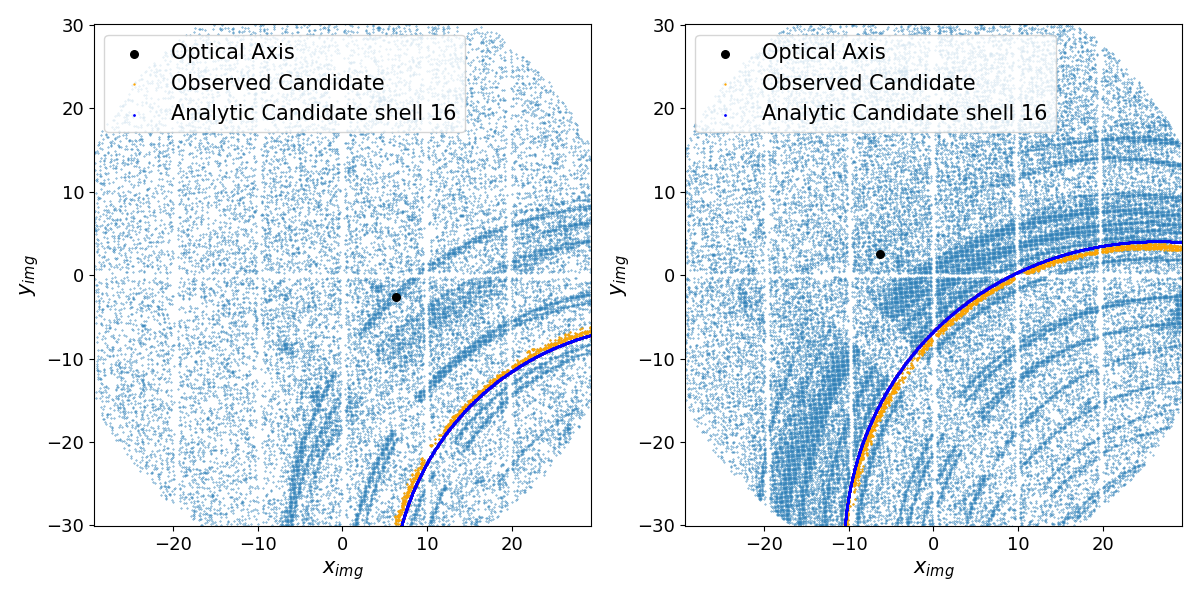}
    \caption{Shell 16 analytically produced (blue arcs) and represented in observations 0720540501 (left) and 0720540601 (right) with no tilts. The association process outputs this shell as a plausible candidate for the two selected arcs (orange arcs).}
    \label{fig:shell42pre}
\end{figure}

The Shell misAlignment Detection for straylight Estimation (SHADE) is an algorithm designed to reproduce the stray light arcs observed in X-ray data obtained with Wolter I telescopes, so as to infer the misalignment parameters of the mirror shells. The pipeline begins by manually extracting an arc from an observation and verifying its possible association with non-tilted simulated arcs. We explore the case of observation 0720540501 and visually identify the coordinates of an arc. It is important to note that this process may inadvertently include photons that do not belong to the arc. We use Eqs. \eqref{eqmis1} and \eqref{eqmis2} and generate simulated arcs for each \textit{XMM-Newton} shell to seek a preliminary correspondence with the observed arc. In accordance with the reference frame described in Appendix \ref{appendix:AppendixA}, we determine the source coordinates relative to the detector center as $\theta = 55.71'$ and $\iota = 5.42$ rad.
At first glance, the correspondence was assigned based on a least squares approach between the simulated arcs and the observed one, yielding a correspondence with shell 16 (blue arc) and the orange arcs represented in Fig. \ref{fig:shell42pre}. This approach is based on the assumption that, in the absence of information on the tilt, an arc is more likely to correspond to the simulated position computed with zero tilt. However, this approach implicitly assumes that possible shell tilts do not alter the arc identification, raising questions about the robustness of this criterion.

\begin{figure}
    \centering
    \includegraphics[scale=0.25]{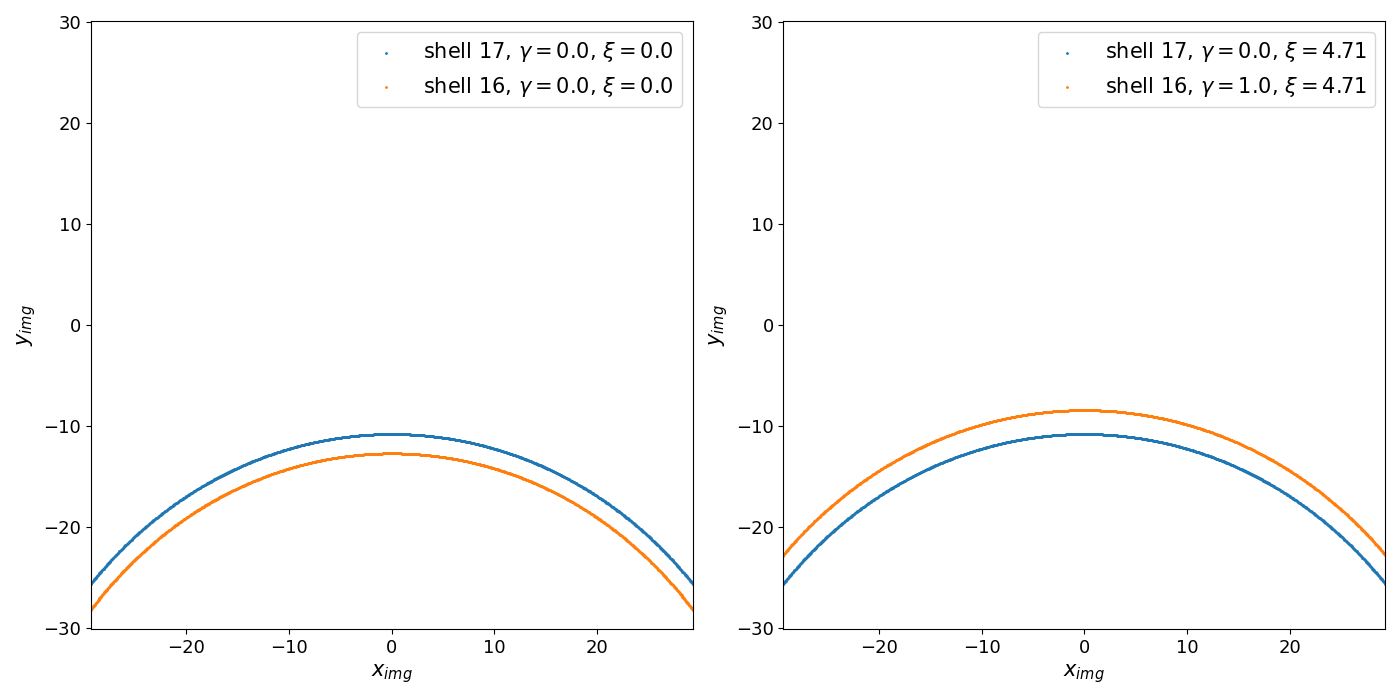}
    \caption{Arc patterns reproduced for shells 16 and 17 in two different configurations. Left: Arcs reproduced under the perfect mirror hypothesis. Right: Configuration where shell 16 is misaligned with amplitudes $\gamma = 1'$ and $\xi = 3\pi/2$. The result of these combinations produces an interchange between the order in which the shells can be visualized on the detector.} 
    \label{fig:changeshell}
\end{figure}

We investigated a configuration involving two neighboring shells to illustrate the potential impact of tilt-induced misalignment on the pattern. We simulated the case of a source with $\theta=1^\circ$ and $\iota=3\pi/2$, generating the arcs from shell 16 and shell 17 using Eqs. \eqref{eqmis1} and \eqref{eqmis2}. In the first case, we assumed $\gamma=0$ for both shells, while in the second case we introduced a tilt with $\gamma=1'$ and $\xi = 3\pi/2$ for shell 16. As shown in Fig. \ref{fig:changeshell}, the tilt displaces shell 16 toward the center of the detector, causing an inversion in the order of the arcs. This result highlights how particular tilt configurations can lead to shell misidentification, representing a potential obstacle for the arc-fitting strategy. This could be mitigated by fitting all shells at the same time; however, this approach is time consuming, and therefore it will be investigated in future studies.

We tested the scenario of misidentification by repeating the association process $N=1500$ times, adopting a uniform distribution of misalignments in the ranges $\gamma \in [0, 1']$ and $\xi \in [0,2\pi]$ as a conservative assumption, allowing for the broadest possible range of tilt values. This choice was motivated by the lack of precise a priori knowledge of the exact distribution of misalignments and aimed to avoid underestimating their potential effect on the arc association process. Under this assumption, we find that $\sim45\%$ of the associations are consistent with the non-tilted case, suggesting that the presence of tilts could play a significant role. However, we note that this estimate is likely a lower limit on the true association rate, as the real distribution of misalignments may favor smaller values. 

The initial assumption gains further support with the inclusion of 0720540601 as a second observation in the association process. We investigated the possibility of associating arcs across multiple observations simultaneously by randomly selecting $\gamma$ and $\xi$, as previously done, keeping only the configurations where the associations referred to the same shell. Repeating the association test with two observations yields a probability of $\sim 97\%$ for shell 16 {(blue arcs in Fig. \ref{fig:shell42pre})} and the orange ones, remaining the best candidate.

\subsection{Optimization and case study}
\begin{figure}
    \centering
    \includegraphics[scale= 0.6]{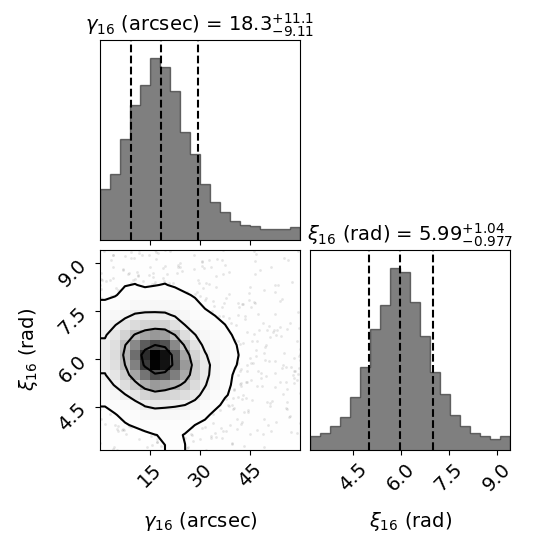}
    \caption{Corner plot of the final distributions for the tilt parameters and the PAs for the 0720540501 and 0720540601 observations. The confidence interval $\sigma$ is estimated by considering the 16\% and 84\% quantiles.}
    \label{fig:fit_dist}
\end{figure}

\begin{figure}
    \centering
    \includegraphics[scale = 0.29]{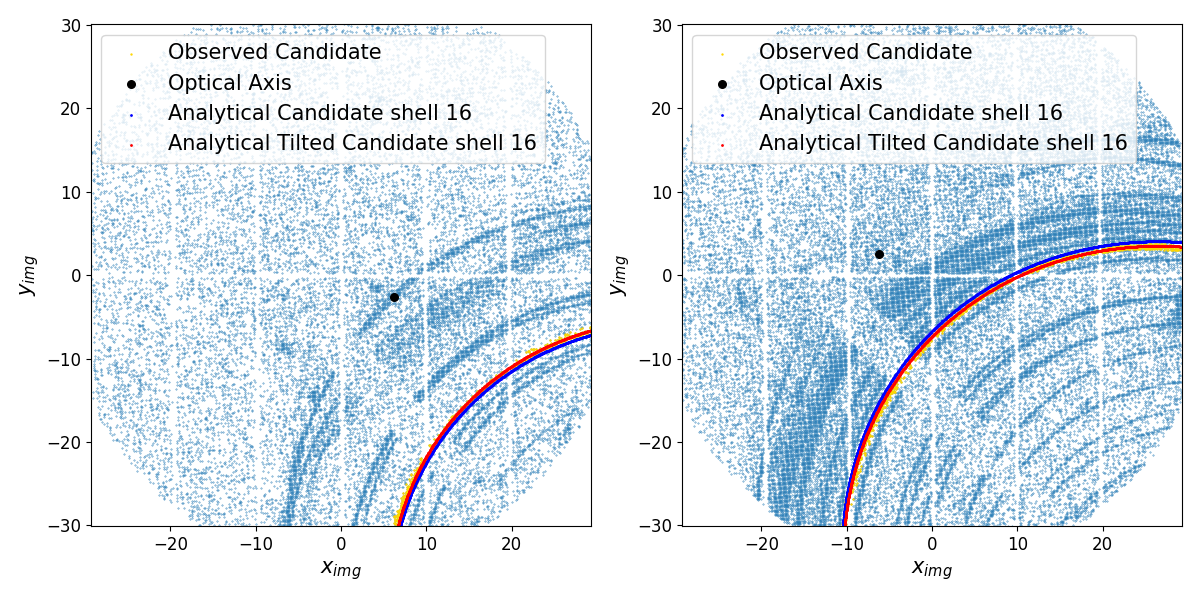}
    \caption{Fit obtained using the TPE algorithm minimizing the sum of the $\chi^2$ metrics calculated for the two observations with a simulated arc for shell 16. The combination of the two-parameter fit can reproduce the arc geometry (red arc), compared with the original produced, which was produced with no tilt (blue arc).}
    \label{fig:final_fit}
\end{figure}

The extraction and simulation procedures were integrated into a general function that outputs a metric corresponding to the sum of the mean squared errors ($\chi^2$) between the observed arcs and the best candidate shells from the association process:
\begin{equation} 
\mu_{\text{F}} = \mu_1 + \mu_2,
\end{equation} 
where $\mu_1$ and $\mu_2$ denote the $\chi^2$ computed for the two observations analyzed. The minimization of this metric was performed using an optimization process based on the Tree-Parzen Estimator (TPE; \citealt{watanabe2023treestructuredparzenestimatorunderstanding}), aimed at finding the best-fit for the tilt parameters.
Observations are aligned with the reference frame aligned with the \textcolor{black}{NCP}, as described in Appendix \ref{appendix:AppendixA}. 
We applied the SHADE algorithm to observations 0720540501 and 0720540601 to reproduce the case of shell 16, respectively with $(\theta_1 = 55.71', \iota_1= 0.84 
\hspace{1mm} \text{rad})$ and $(\theta_2 = 57.78', \iota_2= 3.96 \hspace{1mm} \text{rad})$ based on the considerations from Appendix \ref{appendix:AppendixA}. We excluded $(\delta,\epsilon)$ from the parametrization due to their low impact. The association process identifies two candidate arcs, as shown in Fig. \ref{fig:shell42pre}. The optimization was performed using $N = 2500$ samples, resulting in the parameter distributions shown in Fig. \ref{fig:fit_dist}. We report in Table \ref{table1} the 50\% quantiles represented as $Q_{50}$. Uncertainty estimates were derived from the 16th and 84th percentiles of the parameter distributions, corresponding to the 68\% confidence interval (1$\sigma$). Although the best-fit values do not necessarily coincide with the distribution peaks, they fall well within the 1$\sigma$ regions, and thus the reported uncertainties reflect the expected parameter spread.
To produce the final analytical curves, we generated samples from the distribution of the best fit values output from the pipeline and generated the simulated arc for shell 16 using Eqs. \eqref{eqmis1} and \eqref{eqmis2}. The result obtained is shown in Fig. \ref{fig:final_fit}. The comparison demonstrates that the optimized parameters improve the similarity between the observed and the simulated arc in both observations for this particular shell.

\begin{table}[h!]
\caption{Best-fit values for the tilt parameters}
\label{table1}
\centering
\renewcommand{\arraystretch}{} 
\scalebox{1}{
\begin{tabular}{|c|c|c|}
\hline
Parameter & $Q^{(2)}_{50}$ & $Q^{(3)}_{50}$ \\ 
\hline
\textbf{$\gamma_{16}$} (arcsec)
 & $18.3^{+11.1}_{-9.11}$ &   $21.9^{+10.3}_{-9.02}$ \\
\hline
 \textbf{$\xi_{16}$} (rad) & $5.99^{+1.04}_{-0.98}$ & $5.88^{+1.02}_{-0.97}$ \\
 
\hline
\end{tabular}
}
\newline
\end{table}

\subsection{Testing on more observations}

\begin{figure}
    \centering
    \includegraphics[scale=0.28]{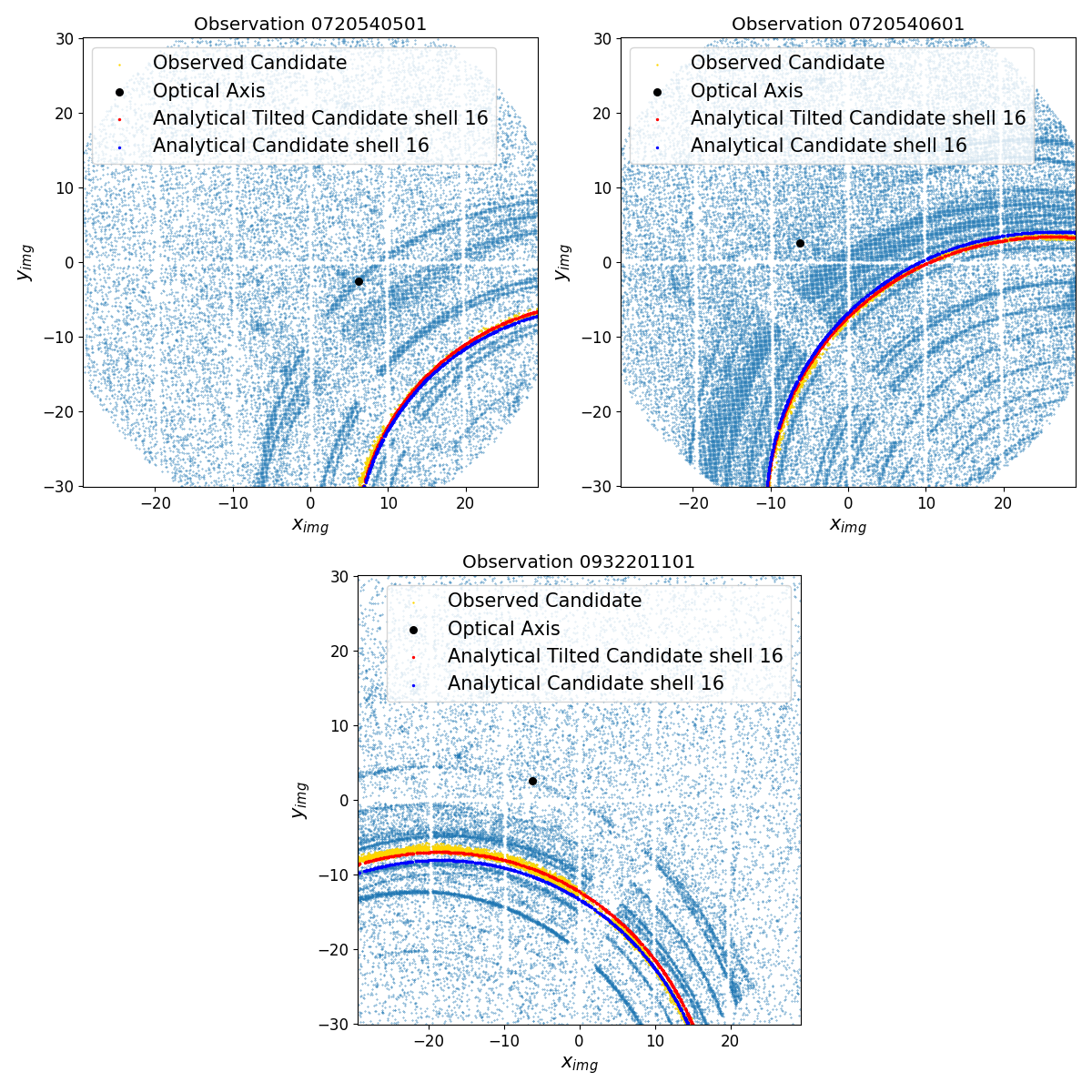}
    \caption{Representation of analytical shell 16 assuming the tilt parameters reported in Table \ref{table1} (red arcs), compared to the same shell produced assuming no tilt (blue arc). The best-fit values improve the similarity of the arc geometry in all three observations.}
    \label{fig:obsgrid2}
\end{figure}

We assessed the reliability of the tilt parameters reported in Table \ref{table1} by testing their consistency with a third observation. In this section, we consider the observation 0932201101, shown in Fig. \ref{fig:0932201101}, as a validation test. This observation was performed around GX5-1 at an angular distance of $\theta = 57.44'$ from the detector origin. The similarity of this off-axis angle with that of the previous observations suggests that the same arc from shell 16 should be detectable. This observation provides a similar PA to 0720540601 but with a different orientation with respect to the source, allowing us to investigate a more general configuration for the algorithm. We extracted the polar coordinates of the source with respect to the optical axis, obtaining $\theta = 57.51'$ and $\iota = 4.97$ rad. Using these coordinates, we generated the simulated arc for shell 16 for all three observations considered, finding a plausible match in 0932201101 (gold arc in the bottom panel in Fig. \ref{fig:obsgrid2}).

We repeated the simulation process mentioned in the previous paragraph to generate shell 16 in all observations, obtaining the result shown in Fig. \ref{fig:obsgrid2}. This result proves that adding this observation increases confidence in the consistency of the arcs, although it cannot guarantee the minimization of all the $\chi^2$ values of the individual observations due to the presence of correlations. Notably, observation 0932201101 was performed in 2024 \citep{ponti2019x}, approximately 24 years after the earliest observation in our sample. This result supports the hypothesis of the temporal stability of the stray light pattern and the underlying shell misalignment.

\begin{figure}
    \centering
    \includegraphics[scale=0.6]{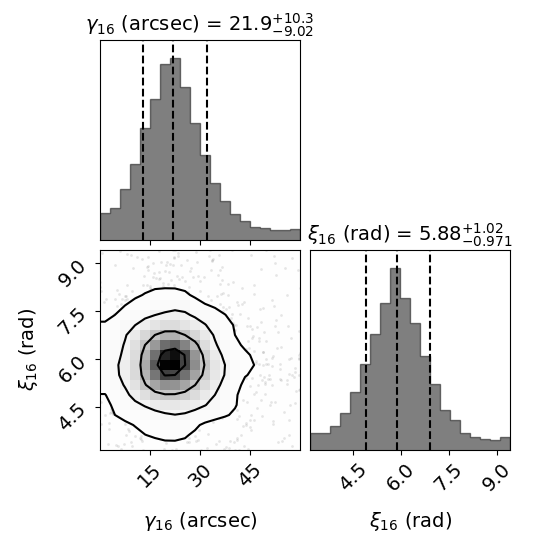}
    \caption{Representation of the distribution of the parameters obtained by the correlated fit performed for three observations. The confidence interval $\sigma$ is estimated by considering the 16\% and 84\% quantiles.}
    \label{fig:fincorner}
\end{figure}

We integrated observation 0932201101 into the SHADE pipeline along with the two previous observations. The optimization outputs the parameter distributions shown in Fig. \ref{fig:fincorner}. We followed the previous procedure, reporting the quantiles in Table \ref{table1}. Including observation 0932201101 provides consistent distributions when comparing Figs. \ref{fig:fit_dist} and \ref{fig:fincorner}. The best-fit parameters were used to produce tilted arcs that align accordingly with the candidates in the observations.

\subsection{Parameters estimation of double Arcs}

\begin{figure}
    \centering
    \includegraphics[scale=0.35]{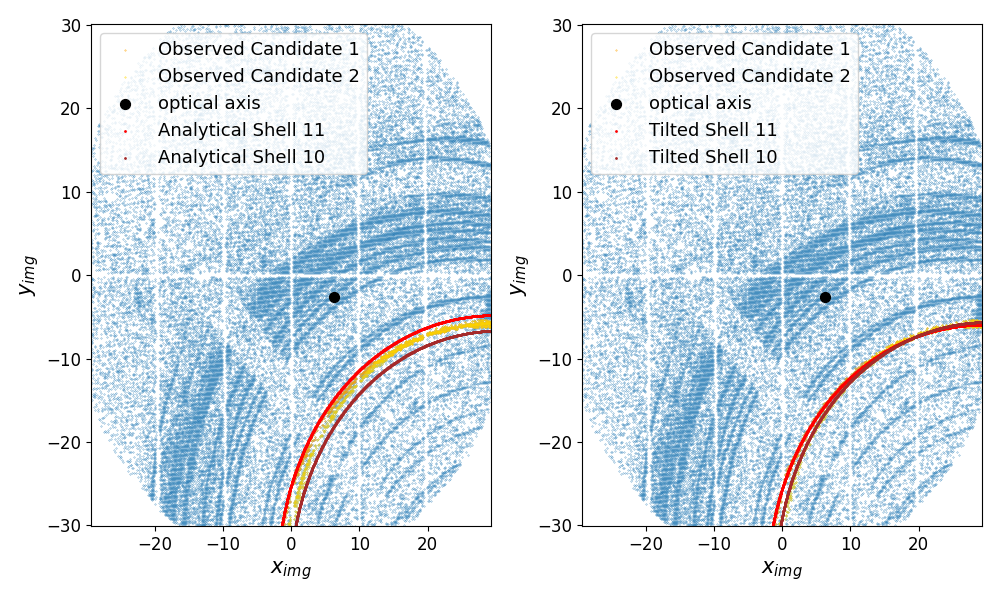}
    \caption{Observation 0720540601 showing analytical arcs for shell 11 and shell 10 without tilts (left panel) and with best-fit parameters extrapolated from the SHADE pipeline (right panel). The golden and orange arcs represent the candidates chosen for parameter estimation.}
    \label{fig:fit_doublearcs}
\end{figure}

\begin{figure}
    \centering
    \includegraphics[scale=0.36]{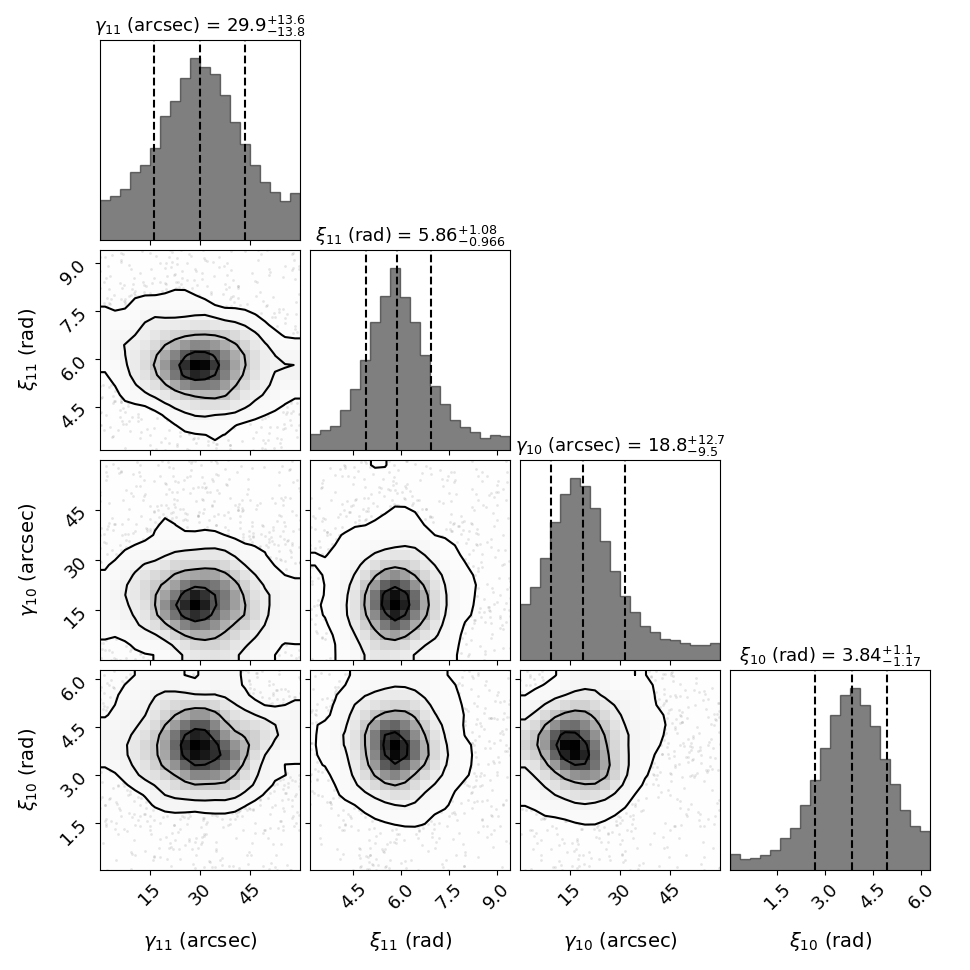}
    \caption{Distributions of the tilts for the double arcs represented in Fig. \ref{fig:fit_doublearcs} for shells 11 and 10.}
    \label{fig:distributions_doublearcs}
\end{figure}

Testing the algorithm to provide tilt parameter measurements of the double arcs required a change in our initial approach. Instead of considering multiple observations, we applied the algorithm to 0720540601 to estimate the parameters of the double arcs. \textcolor{black}{In contrast to} the previous case, we correlated two arcs in the same observation and verified that the algorithm performed the parameter estimation of these arcs simultaneously. This process also \textcolor{black}{hinder} the association since we did not search for counterparts in other observations. Due to the lack of data with feature similar to this observation, extending the parameter estimation to multiple observations has been difficult. Figure \ref{fig:fit_doublearcs} (left panel) shows the two candidates (orange and gold) analyzed and the corresponding analytical match obtained by the same association process described in the previous section. This process output shell 10 and 11 as the best candidates. Using the same number, $N=2500$, of samples to infer the parameters, the results are shown in Fig. \ref{fig:distributions_doublearcs}. The algorithm provides solid distributions for the parameters, with the subscript indicating the shell to which they refer to. Considering the best-fit parameters from these distributions, we reproduced the double arcs constituted by the two candidates in Fig. \ref{fig:fit_doublearcs} (right panel). Table \ref{tab:table3} sums up the values of the quantiles for these two shells.

\begin{table}[h!]
\caption{Best-fit values for the tilt parameters}
\label{tab:table3}
\centering
\renewcommand{\arraystretch}{} 
\scalebox{0.85}{
\begin{tabular}{|c|c|c|c|c|}
\hline
Parameter & $\gamma_{11}$ (arcsec) & $\xi_{11}$ (rad)& $\gamma_{10}$ (arcsec)& $\xi_{10}$ (rad)\\
\hline
$Q_{50}$ & $29.9^{+13.6}_{-13.8}$ & $5.86^{+1.08}_{-0.97}$ & $18.8^{+12.7}_{-9.5}$ & $3.84^{+1.10}_{-1.17}$ \\
\hline
\end{tabular}
}
\newline
\end{table}

\section{Final discussion}

We applied the SHADE algorithm to reproduce the geometry of stray light arcs in \textit{XMM-Newton} observations, focusing mostly on shell 16 as a representative case study. The agreement between the simulated and observed arcs is encouraging; however, several aspects require further investigation. The association process between the shells and the arcs represents a potential source of uncertainty, which could be mitigated by combining information from multiple observations. Additionally, the extraction of the arcs, currently based on visual inspection, would benefit from the development of an automated approach. In this context, machine learning techniques, such as neural networks, could offer a robust solution.
The assumption of time invariance of the stray light pattern is another cornerstone of this work. The consistency observed between data spanning over two decades supports this hypothesis. Nevertheless, future observations targeting the same aimpoint as those considered in this study could further strengthen this conclusion, providing additional evidence for the stability of shell misalignments over time. The possibility of including the deformation effects of the mirrors could provide more information and better agreement between the observed and the simulated arcs.
Future developments of SHADE could also involve a more comprehensive parameterization, accounting for additional systematic effects, such as shifts of the baffles, as explored in \cite{Friedrich}. This aspect was not included in the present work to maintain a simpler description of the parameterization. However, its contribution could be crucial for further improving the reliability of the tilt measurements and gaining a more complete understanding of the mirror module alignment.

\section{Conclusions}
In this paper, we presented SHADE (The Shell misAlignment Detection for Straylight Estimation), a novel algorithm designed to reproduce the stray light arcs in Wolter I telescopes by inferring the misalignment parameters of individual mirror shells. This approach enhances the consistency between observed stray light patterns and simulations, enabling better predictions for future observations. We applied SHADE to \textit{XMM-Newton} data, focusing on shell 16, and obtained the tilt parameters presented in Table \ref{table1} by combining three observations. As a proof of concept, we tested the algorithm on double arcs belonging to the same observation, obtaining results reported in Table \ref{tab:table3} for shell 10 and 11. These results demonstrate the capability of SHADE to provide quantitative estimates of shell tilts, offering a pathway to refine the modeling of stray light patterns and improve the calibration of future X-ray observations.

\begin{acknowledgements}
We thank P. Friedrich for the helpful discussions. We also thank D. Parodi and G. Valsecchi for the precious insights.
This work is based on observations obtained with \textit{XMM-Newton}, an ESA science mission with instruments and contributions directly funded by ESA Member States and NASA. This project acknowledges financial support from the European Research Council (ERC) under the European Union’s Horizon 2020 research and innovation program HotMilk (grant agreement No.
865637), support from Bando per il Finanziamento della Ricerca Fondamentale 2022 dell’Istituto Nazionale di Astrofisica (INAF): GO Large program and from the Framework per l’Attrazione e il Rafforzamento delle Eccellenze (FARE) per la ricerca in Italia (R20L5S39T9).
\end{acknowledgements}

\newpage
\bibliographystyle{aa} 
\bibliography{bibliography}

\begin{thebibliography}{39}
\expandafter\ifx\csname natexlab\endcsname\relax\def\natexlab#1{#1}\fi

\bibitem[{Aschenbach(2002)}]{aschenbach2002orbit}
Aschenbach, B. 2002, in X-Ray Optics for Astronomy: Telescopes, Multilayers,
  Spectrometers, and Missions, Vol. 4496, SPIE, 8--22

\bibitem[{Bhulla {et~al.}(2019)Bhulla, Misra, Yadav, \&
  Jaaffrey}]{bhulla2019astrosat}
Bhulla, Y., Misra, R., Yadav, J., \& Jaaffrey, S. 2019, RAA, 19, 114

\bibitem[{Conconi \& Campana(2001)}]{conconi2001optimization}
Conconi, P. \& Campana, S. 2001, A\&A, 372, 1088

\bibitem[{de~Chambure {et~al.}(1999{\natexlab{a}})de~Chambure, Lain{\'e},
  Van~Katwijk, \& Kletzkine}]{de1999xmm}
de~Chambure, D., Lain{\'e}, R., Van~Katwijk, K., \& Kletzkine, P.
  1999{\natexlab{a}}, ESA Bulletin, 100, 30

\bibitem[{de~Chambure {et~al.}(1999{\natexlab{b}})de~Chambure, Laine, van
  Katwijk, Ruehe, Schink, Hoelzle, Gutierrez, Domingo, Ibarretxe, Tock,
  {et~al.}}]{de1999x}
de~Chambure, D., Laine, R., van Katwijk, K., {et~al.} 1999{\natexlab{b}}, in
  Design and Engineering of Optical Systems II, Vol. 3737, SPIE, 396--408

\bibitem[{de~Chambure {et~al.}(1997)de~Chambure, Lain{\'e}, vanKatwijk,
  vanCasteren, \& Glaude}]{dechambure1997producing}
de~Chambure, D., Lain{\'e}, R., vanKatwijk, K., vanCasteren, J., \& Glaude, P.
  1997, ESA Bulletin, 68

\bibitem[{ESA(2011)}]{RefSys}
ESA. 2011, Spatial coordinate systems,
  \url{https://xmm-tools.cosmos.esa.int/external/xmm_calibration/calib/documentation/CALHB/node7.html}

\bibitem[{ESA(2023)}]{Boresight}
ESA. 2023, epicbscalgen,
  \url{https://xmm-tools.cosmos.esa.int/external/sas/current/doc/epicbscalgen.pdf}

\bibitem[{ESA(2025)}]{ecoordconv}
ESA. 2025,
  \url{https://xmm-tools.cosmos.esa.int/external/sas/current/doc/ecoordconv.pdf}

\bibitem[{Freyberg(2006)}]{freyberg2006}
Freyberg, M. 2006, XMM-Newton XRT+EPIC straylight: overview,
  \url{https://www.cosmos.esa.int/documents/332006/1301262/mjf.pdf}

\bibitem[{Friedrich \& Freyberg(2024)}]{Friedrich}
Friedrich, P. \& Freyberg, M. 2024, Investigations of the XMM-Newton X-ray
  baffle alignment,
  \url{https://axro.cz/wp-content/uploads/conference_uploads/2024/contributions/posters/peter_friedrich_investigations_of_the_xmm-newt_2024.pdf}

\bibitem[{Glatzel {et~al.}(1994)Glatzel, Schmidt, Egle, Pauschinger, Schulz,
  Burkert, \& Braeuninger}]{glatzel1994assembly}
Glatzel, H., Schmidt, M., Egle, W.~J., {et~al.} 1994, in Space Optics 1994:
  Space Instrumentation and Spacecraft Optics, Vol. 2210, SPIE, 360--372

\bibitem[{Jansen {et~al.}(2001)Jansen, Lumb, Altieri, Clavel, Ehle, Erd,
  Gabriel, Guainazzi, Gondoin, Much, {et~al.}}]{jansen2001xmm}
Jansen, F., Lumb, D., Altieri, B., {et~al.} 2001, A\&A, 365, L1

\bibitem[{Kirsch(2004)}]{OptAxis}
Kirsch, M. 2004, Improved Vignetting Corretion by refining the XMM optial
  axis., \url{https://xmmweb.esac.esa.int/docs/documents/CAL-SRN-0156-1-3.pdf}

\bibitem[{Marioni {et~al.}(1999)Marioni, Radaelli, Raggio, de~Chambure, \&
  Laine}]{marioni1999xmm}
Marioni, F., Radaelli, P.~G., Raggio, M.~E., de~Chambure, D., \& Laine, R.
  1999, in Optical Fabrication and Testing, Vol. 3739, SPIE, 232--244

\bibitem[{Mitsuda {et~al.}(1985)Mitsuda, Inoue, Koyama, Makishima, Matsuoka,
  Ogawara, Shibazaki, Suzuki, Tanaka, \& Hirano}]{mitsuda1985energy}
Mitsuda, K., Inoue, H., Koyama, K., {et~al.} 1985, PASJ, 36, 741

\bibitem[{Mori \& Friedrich(2022)}]{mori2022collimators}
Mori, H. \& Friedrich, P. 2022, in Handbook of X-ray and Gamma-ray Astrophysics
  (Springer), 1--44

\bibitem[{Mu{\~n}oz-Darias {et~al.}(2014)Mu{\~n}oz-Darias, Fender, Motta, \&
  Belloni}]{munoz2014black}
Mu{\~n}oz-Darias, T., Fender, R., Motta, S., \& Belloni, T. 2014, MNRAS, 443,
  3270

\bibitem[{Pantaleoni {et~al.}(2012)Pantaleoni, Kirsch, Martin, McDonald, \&
  Tuttlebee}]{pantaleoni2010xmm}
Pantaleoni, M., Kirsch, M., Martin, J., McDonald, A., \& Tuttlebee, M. 2012, in
  SpaceOps 2010 Conference Delivering on the Dream Hosted by NASA Marshall
  Space Flight Center and Organized by AIAA, 1980

\bibitem[{Ponti {et~al.}(2019)Ponti, Hofmann, Churazov, Morris, Haberl, Nandra,
  Terrier, Clavel, \& Goldwurm}]{ponti2019x}
Ponti, G., Hofmann, F., Churazov, E., {et~al.} 2019, Nature, 567, 347

\bibitem[{Ponti {et~al.}(2015)Ponti, Morris, Terrier, Haberl, Sturm, Clavel,
  Soldi, Goldwurm, Predehl, Nandra, {et~al.}}]{ponti2015}
Ponti, G., Morris, M., Terrier, R., {et~al.} 2015, MNRAS, 453, 172

\bibitem[{Schartel {et~al.}(2024)Schartel, Gonz{\'a}lez-Riestra, Kretschmar,
  Kirsch, Rodr{\'\i}guez-Pascual, Rosen, Santos-Lle{\'o}, Smith, Stuhlinger, \&
  Verdugo-Rodrigo}]{schartel2024xmm}
Schartel, N., Gonz{\'a}lez-Riestra, R., Kretschmar, P., {et~al.} 2024, in
  Handbook of X-ray and Gamma-ray Astrophysics (Springer), 1501--1538

\bibitem[{Simonyan \&
  Zisserman(2015)}]{simonyan2015deepconvolutionalnetworkslargescale}
Simonyan, K. \& Zisserman, A. 2015, Very Deep Convolutional Networks for
  Large-Scale Image Recognition

\bibitem[{SOC(2024)}]{xmmhu}
SOC, E. X.-N. 2024, ”XMM-Newton Users Handbook”, issue 2.22

\bibitem[{Spiga(2011)}]{spiga2011optics}
Spiga, D. 2011, A\&A, 529, A18

\bibitem[{Spiga(2015)}]{spiga2015analytical}
Spiga, D. 2015, in Optics for EUV, X-Ray, and Gamma-Ray Astronomy VII, Vol.
  9603, SPIE, 107--124

\bibitem[{Spiga(2016)}]{spiga2016x}
Spiga, D. 2016, in Space Telescopes and Instrumentation 2016: Ultraviolet to
  Gamma Ray, Vol. 9905, SPIE, 1951--1961

\bibitem[{Spiga {et~al.}(2009)Spiga, Cotroneo, Basso, \&
  Conconi}]{spiga2009analytical}
Spiga, D., Cotroneo, V., Basso, S., \& Conconi, P. 2009, A\&A, 505, 373

\bibitem[{Stockman {et~al.}(1999)Stockman, Barzin, Hansen, Mazy, Tock,
  de~Chambure, Laine, Kampf, Banham, Canali, {et~al.}}]{stockman1999xmm}
Stockman, Y., Barzin, P., Hansen, H., {et~al.} 1999, in X-Ray Optics,
  Instruments, and Missions II, Vol. 3766, SPIE, 51--61

\bibitem[{Stramaccioni {et~al.}(2000)Stramaccioni, Faust, \&
  Hinger}]{stramaccioni2000xmm}
Stramaccioni, D., Faust, T., \& Hinger, J. 2000, SAE transactions, 499

\bibitem[{Str{\"u}der {et~al.}(2001)Str{\"u}der, Briel, Dennerl, Hartmann,
  Kendziorra, Meidinger, Pfeffermann, Reppin, Aschenbach, Bornemann,
  {et~al.}}]{struder2001european}
Str{\"u}der, L., Briel, U., Dennerl, K., {et~al.} 2001, A\&A, 365, L18

\bibitem[{T. {et~al.}(2003)T., J., A., A., colleagues at MPE~in Munich, \&
  Center}]{damage}
T., A., J., C., A., R., {et~al.} 2003,
  \url{https://www.cosmos.esa.int/documents/332006/1157523/175772_afa_micrometeoroids.pdf}

\bibitem[{Tanaka(1997)}]{tanaka1997x}
Tanaka, Y. 1997, in Accretion Disks—New Aspects: Proceedings of the EARA
  Workshop Held in Garching, Germany, 21--23 October 1996, Springer, 1--20

\bibitem[{Virtanen {et~al.}(2021)Virtanen, Gommers, Burovski, Oliphant,
  Weckesser, Cournapeau, Peterson, Reddy, Haberland, Wilson,
  {et~al.}}]{virtanen2021scipy}
Virtanen, P., Gommers, R., Burovski, E., {et~al.} 2021, Zenodo

\bibitem[{Virtanen {et~al.}(2020)Virtanen, Gommers, Oliphant, Haberland, Reddy,
  Cournapeau, Burovski, Peterson, Weckesser, Bright, {van der Walt}, Brett,
  Wilson, Millman, Mayorov, Nelson, Jones, Kern, Larson, Carey, Polat, Feng,
  Moore, {VanderPlas}, Laxalde, Perktold, Cimrman, Henriksen, Quintero, Harris,
  Archibald, Ribeiro, Pedregosa, {van Mulbregt}, \& {SciPy 1.0
  Contributors}}]{2020SciPy-NMeth}
Virtanen, P., Gommers, R., Oliphant, T.~E., {et~al.} 2020, Nat. Methods, 17,
  261

\bibitem[{Wasserstein \& Lazar(2016)}]{wasserstein2016asa}
Wasserstein, R.~L. \& Lazar, N.~A. 2016, The ASA statement on p-values:
  context, process, and purpose

\bibitem[{Watanabe(2023)}]{watanabe2023treestructuredparzenestimatorunderstanding}
Watanabe, S. 2023, Tree-Structured Parzen Estimator: Understanding Its
  Algorithm Components and Their Roles for Better Empirical Performance

\bibitem[{Wolter(1952)}]{wolter1952spiegelsysteme}
Wolter, H. 1952, Ann. Phys., 445, 94

\bibitem[{Y. {et~al.}(2001)Y., P., I., P., H., P., A., \&
  G.}]{stockman2001environmental}
Y., S., P., B., I., D., {et~al.} 2001, in Fourth International Symposium
  Environmental Testing for Space Programmes, Vol. 467, 11

\end{thebibliography}

\newpage

\begin{appendix}

\section{XMM features} 
\label{appendix:AppendixA}
Appendix A presents technical and physical information about \textit{XMM-Newton}, focusing on the features that have the greatest impact on the stray light arcs. 

\subsection{Mirror modules}
\begin{figure}
    \centering
    \includegraphics[scale = 0.45]{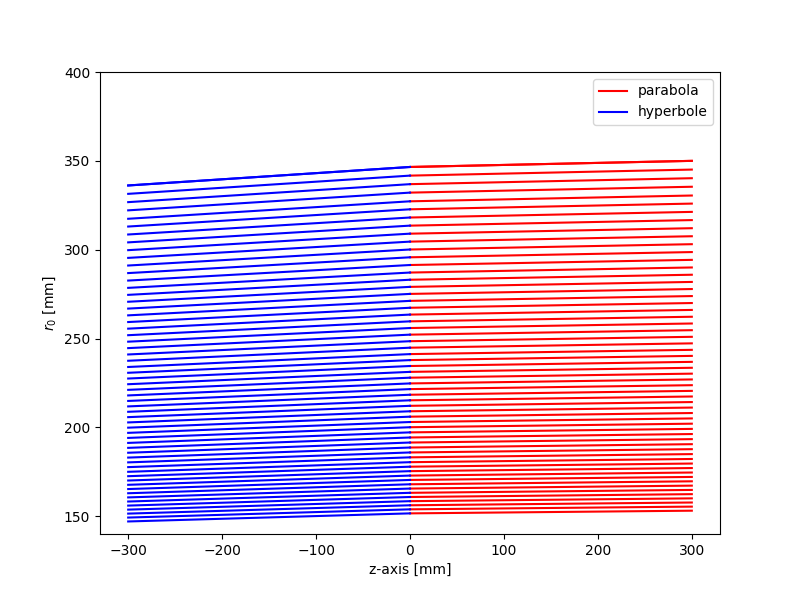}
    \caption{Shell profile of the \textit{XMM-Newton} Observatory. The different slope and distance from the optical axis output the focus of the photons on the detector at a certain focal length.}
    \label{fig:mirrormodule}
\end{figure}

\textit{XMM-Newton} presents three mirror modules constituted by 58 shells formed by a parabola and an hyperbola. The purpose of this configuration is to focus the double reflected light rays which impinge on the shells on a detector at distance $F=7500$ mm. This distance represents the focal length of the mirror module \citep{schartel2024xmm}. The shells can be mathematically modeled by parametric equations \citep{conconi2001optimization}. Both parabola and hyperbola have $L=300$ mm length. Their inclination forms two angles with the z-axis which is perpendicular to the detector and coincides with the optical axis. The angle between the parabola and the z-axis is named $\alpha_0$ and it differs among shells. This angle depends on the distance, called $R_0$, between the shell and the z-axis calculated at $z=0$. The plane $z=0$ is the IP. For a whole review of the physical features of the telescope we refer to \cite{jansen2001xmm} and \cite{schartel2024xmm}. The greater $R_0$ is, the higher is the slope. This configuration allows to reproduce the focusing process we anticipated. The hyperbola forms an angle $\beta_0 \sim 3\alpha_0$. A more detailed description of the geometry of the mirror module is reported in \cite{spiga2015analytical} and \cite{stramaccioni2000xmm}. Figure \ref{fig:mirrormodule} depicts the mirror profiles of the 58 shells involved.

\subsection{Baffles}

The baffle represents a structure added to remove stray light from observations. \textit{XMM-Newton} presents a baffle system constituted by two cylindrical structures, mounted on each of the 58 shells. The cylinders have been separated into 16 slices due to the presence of a spider, which serves as a support structure. The 58 baffles have different heights and are supposed to be perfectly aligned with the mirrors. The effect of such structure is to obscure the mirrors from light rays from off-axis sources at $\theta>1.4$ degrees \citep{de1999x}. A more comprehensive description of the baffle structure can be found in \cite{de1999x}, \cite{aschenbach2002orbit} and \cite{mori2022collimators}. 
It has been demonstrated that, despite the presence of the baffle, the effective area of stray light for \textit{XMM-Newton} is, respectively, 1\% at 2 keV and 1.4\% at 8 keV, in comparison to the area of the double reflections. This indicates that an off-axis source 100 times more luminous than the on-axis target may generate count rates in ghost images that are comparable to the target rate, as shown in \cite{de1999x}. For that reason, the addition of the baffle does not ensure a definitive solution to the stray light problem.

\begin{figure}
    \centering
    \includegraphics[scale=0.5]{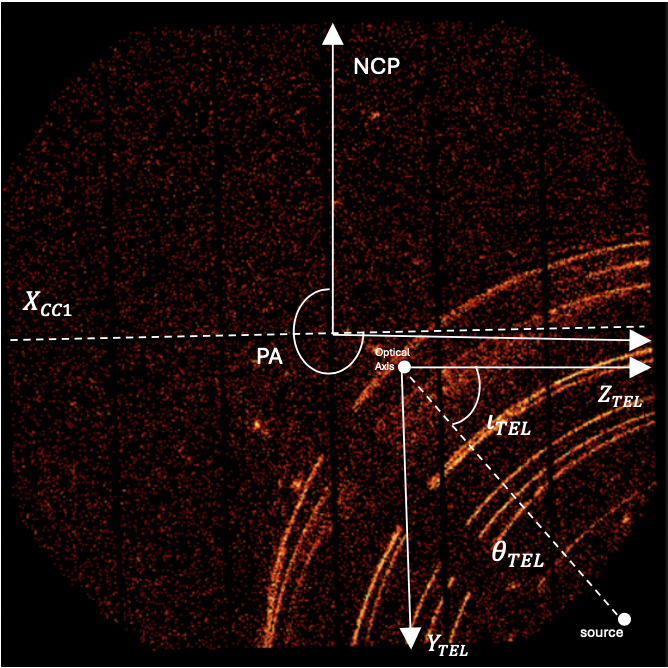}
    \caption{Reference frame configuration considered in this work for observation 0720540501, represented in its mirrored form. The task "ecoordconv" provides coordinates $(\theta_{\text{TEL}}, \iota_{\text{TEL}})$ in the TelCoord frame, which is misaligned with respect to the \textcolor{black}{NCP} frame at an angle related to the PA. This reference frame is also misaligned with respect to the direction of the x-axis of the detector frame CAMCOORD1, represented by the dashed white line, whose center is placed in the gap of the CCDs.
    Computing the correct angle $\iota$ in the TelCoord frame, the angular difference between the PA and the $x$-axis of the \textcolor{black}{NCP} is taken into account. Since our purpose is to represent the optics in the TelCoord frame, we correct the pattern by accounting for the offset between the optical axis and the geometric center of the detector.}
    \label{fig:RefFrames}
\end{figure}

\subsection{The PN camera frame of reference}

Setting up the stray light pattern simulation for EPIC-pn requires selecting the appropriate reference frame to accurately reproduce the optics. The goal is to represent the stray light pattern in a reference frame with the origin in the geometrical center of the detector and with the $y-axis$ aligned with the \textcolor{black}{NCP}. At first, this purpose implies to account for the presence of the offset between the geometrical center of the detector and the optical axis of the camera \citep{OptAxis}, that represents the reference for the optics of the phenomenon. Combining the optical axis location from calibration data with the pixel coordinates of the detector center, we estimate this offset to be approximately $2'$.
The application of Eqs. \eqref{params_eq_1} and \eqref{params_eq_2} requires the source coordinates in a reference frame centered in the optical axis. This is represented by the TelCoord frame \citep{RefSys}, the origins of which can be projected into the focal plane considering mirroring effects. The task \textit{ecoordconv} \citep{ecoordconv} returns the polar coordinates of a source, defined as $\theta_{\text{TEL}}$, the angular distance from the optical axis, and $\iota_{\text{TEL}}$, the polar angle on the detector plane.
Figure \ref{fig:RefFrames} shows a representation of the reference frames. Normally the detector is also rotated with respect to the \textcolor{black}{NCP} frame by an angle related to the PA , defined as the angle between the \textcolor{black}{NCP} and the detector x-axis (CAMCOORD1) \citep{jansen2001xmm, schartel2024xmm, pantaleoni2010xmm}, which is supposed to be aligned with the CAMCOORD1 $x$-axis as showed in Fig. \ref{fig:RefFrames}. In this paper we assume that TelCoord is oriented according to the PA defined by the XMM Science Archive, which refers to the optical axis reference frame. This assumption implies an additional angular discrepancy between TelCoord and CAMCOORD1. These rotations implies a correction for the image and the coordinates, depending on an empirical value for the PA, calculated for the detector with the same SAS task \textit{ecoordconv}, and the values taken from the XMM Science Archive for the Telcoord rotation. The coordinates considered in this work are expressed by the following equations:
\begin{gather}
    \theta = \theta_{\text{TEL}} \\ 
    \iota = \iota_{\text{TEL}} + (\text{PA}_{\text{ARCH}} - \text{PA}_{\text{EMP}})
\end{gather}
where we stress that the polar angle $\iota_{\text{TEL}}$ is referred to a reference system that rotates with the telescope when the PA changes.
We highlight that Fig. \ref{fig:RefFrames} is a mirrored representation of the observation. 

The offset between the detector center and the optical axis affects the region of the detector illuminated by stray light, depending on the detector orientation in the focal plane. This orientation is again determined by the PA. To account for the offset, we translate the simulated arcs by the opposite displacement, calculated in pixel coordinates.  

Finally, the source coordinates generated by \textit{ecoordconv} also account for the boresight correction. This correction represents the misalignment between the instrument boresight and the satellite coordinate frame. It is described by a set of Euler angles available in ESA calibration files \citep{Boresight}. We estimate that the boresight correction introduces a relative uncertainty of approximately 0.02–0.03\% on both $\theta$ and $\iota$.

The representation of the observations requires a transformation from matrices to coordinates that involve the size of the PN detector as provided by ESA \citep{struder2001european,xmmhu}.

\section{Considerations on ray tracing} 
\label{appendix:AppendixB}

\subsection{Ray tracing}
\begin{figure}
    \centering
    \includegraphics[scale=0.4]{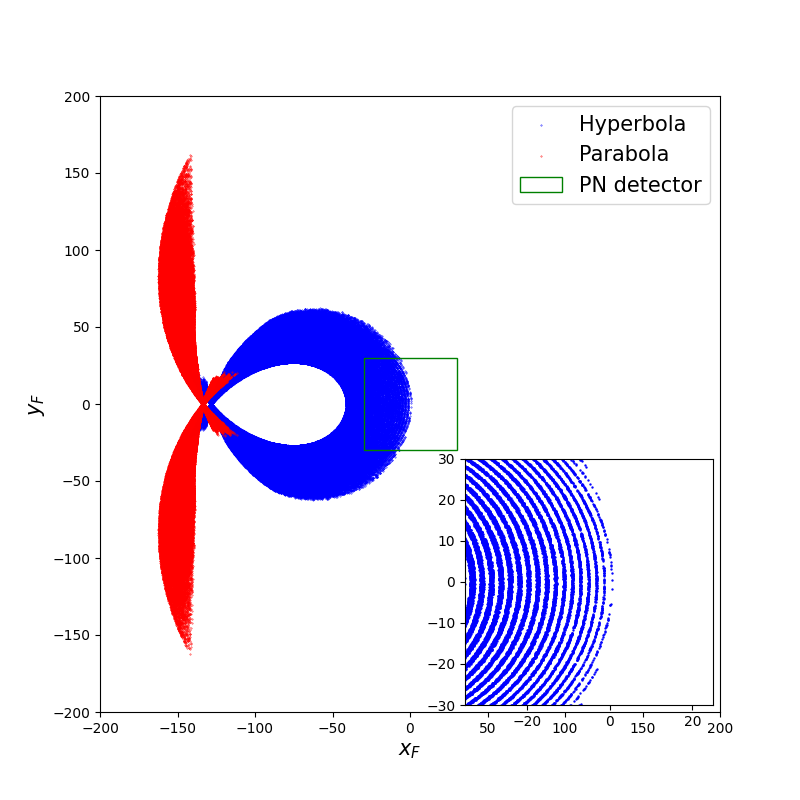}
    \caption{Representation of the stray light pattern produced by both the parabola and the hyperbola with a ray-tracing routine considering $N=10^5$ light rays for each shell.
From this image, it is clear that the pattern features some important characteristics. It is easy to distinguish the impact of the reflections on the parabola and the hyperbola. Singly reflected rays by the parabola (red dots) generate a pattern, which does not intersect the area of the detector. Therefore, this pattern is not included in our analysis. Rays reflected by the hyperbola hit the detector, creating an arched pattern.}
    \label{fig:raytracing}
\end{figure}

\begin{figure}
    \centering
    \includegraphics[scale=0.55]{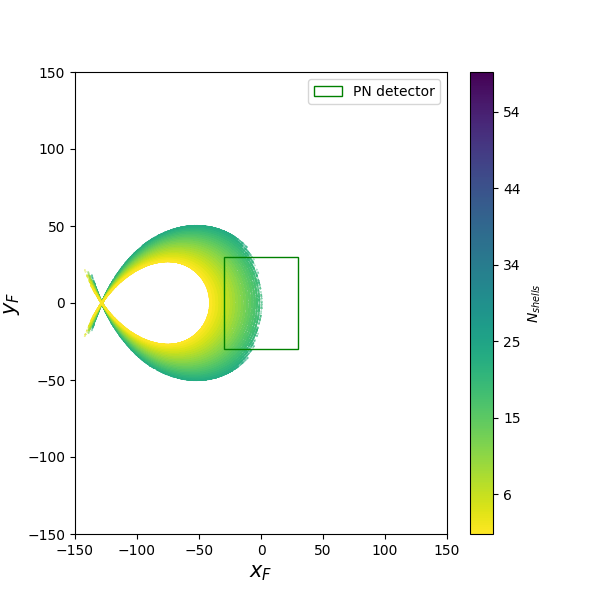}
    \caption{Correspondence between the position of the arc and the shell. Shells with larger $R_0$ are responsible for the arcs along the negative direction, $-x,$ while shells with lower radii produce arcs that do not intersect the detector. The coordinates of this plot refer to the (x,y) focal plane in millimeters.}
    \label{fig:colored shells}
\end{figure}

\begin{figure}
    \centering
    \includegraphics[scale=0.40]{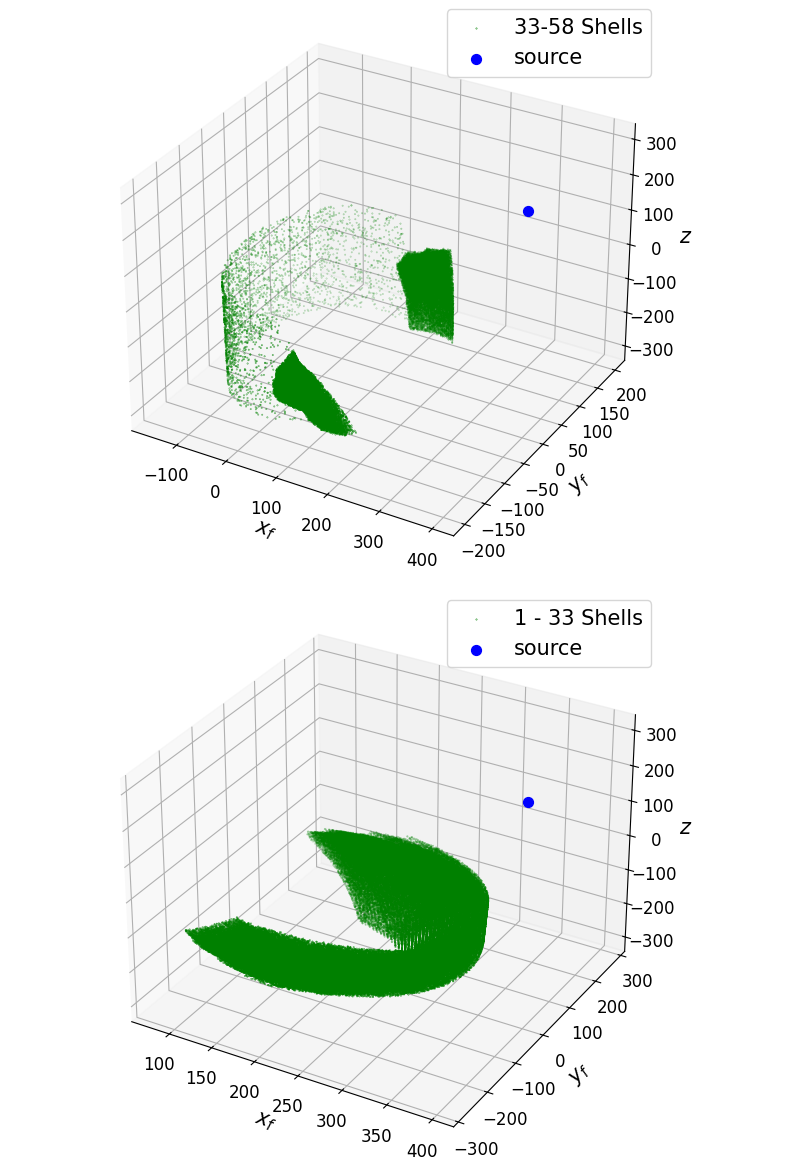}
    \caption{Reflection points of single hyperbola photons for 33-58 shells (upper figure) and shells 1 to 33 (lower figure). These figures are crucial for estimating the contribution of the disposition of the shells to the stray light observed in the detector.}
    \label{fig:hyperbhit}
\end{figure}

Ray tracing can highlight some useful considerations for the stray light arcs, since a range of features observed in \textit{XMM-Newton} images can be explained by the geometrical structure of the mirror module and could also be reproduced by Eqs. \eqref{params_eq_1} and \eqref{params_eq_2}, for example, incomplete arcs that can be modeled accounting for vignetting coefficients \citep{spiga2015analytical}. However, this has not been carried out here.
The ray tracing routine is based on the generation of rays whose path is tracked to see the effect on the detector. We first model the shells of the telescope as a parabola and a hyperbola following \cite{conconi2001optimization} as previously anticipated. Each ray is generated at a random position on the aperture plane of the telescope at a random polar angle $\psi$. Since the rays are supposed to be generated from an off-axis source at a finite distance, we assume they all have the same direction while the velocity components are dictated by $\theta$,
\begin{gather*}
    v_x = -c \cdot \cos \psi \sin \theta \\
    v_y = -c \cdot \sin \psi \sin \theta \\
    v_z = -c \cdot \cos \theta  
\end{gather*}
where $c$ stands for the speed of light and can be set to 1. The negative sign introduced in the equations is necessary to ensure that ray travel from top to bottom, where $z$ is negative. We run the ray tracing routine for $N=10^5$ rays for each shell, assuming a source at distance $\theta = 1'$ and $\iota = 2\pi$. The result of the simulation can be observed in Fig. \ref{fig:raytracing}. In this plot we show the pattern captured by the detector placed in the center and visualized on the focal plane at $z=-f$. It is noticeable that the area of the detector does not encompass the whole stray light pattern but just a limited number of arcs. Figure \ref{fig:raytracing} shows that the latest arcs are not completed. This phenomenon is generated by the regions where the rays are reflected on different shells. We group the interactions with the same shell and obtain Fig. \ref{fig:colored shells} showing that there is a correspondence between the most narrow arcs (yellows) and the the shells with higher $R_0$ and viceversa. In this figure, it is also evident that the 25 most internal shells do not fully form on the detector, resulting in incomplete arcs. This pattern can be linked to the regions of the shells where rays are reflected, as illustrated in Fig. \ref{fig:hyperbhit}. To clarify this behavior, we plot the positions of the reflected light rays from shells 33 to 58 and from shells 1 to 33 separately. The results show that, in the first group, the rays predominantly strike the regions of the hyperbola at 90 degrees from the source direction and, to a minor extent, the side opposite to that of the source. Conversely, in the second group, they tend to hit the part of the hyperbola at the same direction from the source. Light rays have been represented in the 3D space occupied by the hyperbola where $z=0$ and $z=-300$ mm  represent the physical limits of the shells. The source in this plot (blue dot) is only a representative marker to indicate the direction of the rays, whereas in a realistic scenario, the $z$-coordinate corresponds to the actual distance of the source. From these considerations it is clear that only the light rays that interact with a certain interval of shells can reach the detector and participate to the pattern seen in the observations for a given pair of $\theta$ and $\iota$. 
In our example in Fig. \ref{fig:raytracing}, this interval is represented by the shells among the 8th and the 23th.
 
\begin{figure}
    \centering
    \includegraphics[scale=0.5]{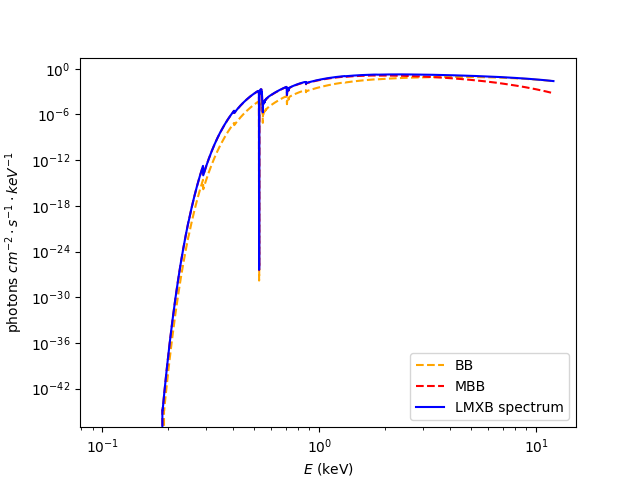}
    \caption{Power spectrum of the LMXB GX5-1. This spectrum (blue line) is the result of the sum of a hard component represented by a black body (dashed black line) and a soft component represented by a multicolor black body (dashed red line). The spectrum also takes into account an absorption coefficient depending on the $N_H$ density at $D=4.7$ kpc.}
    \label{fig:gx5-1 spectrum}
\end{figure}

\subsection{The source spectra}
The observations considered in this work focus on the low-mass X-ray binary (LMXB) GX5-1. The application of ray tracing routines requires consideration of the energy-dependent interaction between photons and mirrors. The emission from LMXBs is well described by a spectral model consisting of two components: a blackbody with temperature $T_{BB} > 2.3$ keV and a multi-color blackbody with an inner disk temperature $T_{MB} = 1.5$ keV, as proposed by \cite{tanaka1997x}. The soft component of the spectrum arises from the accretion disk surrounding the compact object \citep{mitsuda1985energy, munoz2014black}, with $T_{MB}$ representing the temperature at the innermost region of the disk, assumed to be located at $r_{in} \sim 30$ km from the compact object \citep{mitsuda1985energy}. GX5-1 is located at a distance of approximately 4.7 kpc \citep{bhulla2019astrosat}. The total spectrum is further modified by interstellar absorption, which mainly affects the soft X-ray band, and is parameterized by a hydrogen column density of $N_H \sim 10^{22} \text{cm}^{-2}$ \citep{mitsuda1985energy, bhulla2019astrosat}. Figure \ref{fig:gx5-1 spectrum} illustrates the resulting spectrum of GX5-1.

\begin{figure}
    \centering
    \includegraphics[scale=0.45]{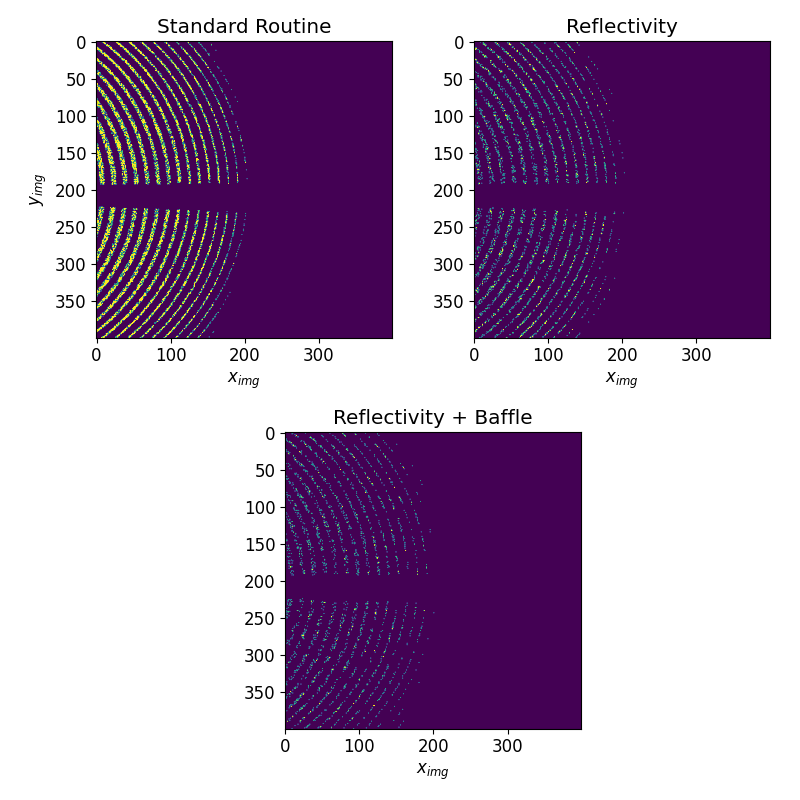}
    \caption{Visual comparison of the three simulations performed with ray tracing. \textcolor{black}{Top - Left}: Results of the routine without considering the contributions of reflectivity and the baffle. \textcolor{black}{Top-right}: Pattern with the addition of reflectivity. \textcolor{black}{Bottom - center}: Contribution of the baffle and reflectivity. It is clear that the geometry of the pattern does not change with the addition of these two contributions but strongly depends on the geometry of the mirror module.}
    \label{fig:fluxcomp}
\end{figure}

\begin{figure}
    \centering
    \includegraphics[scale=0.22]
    {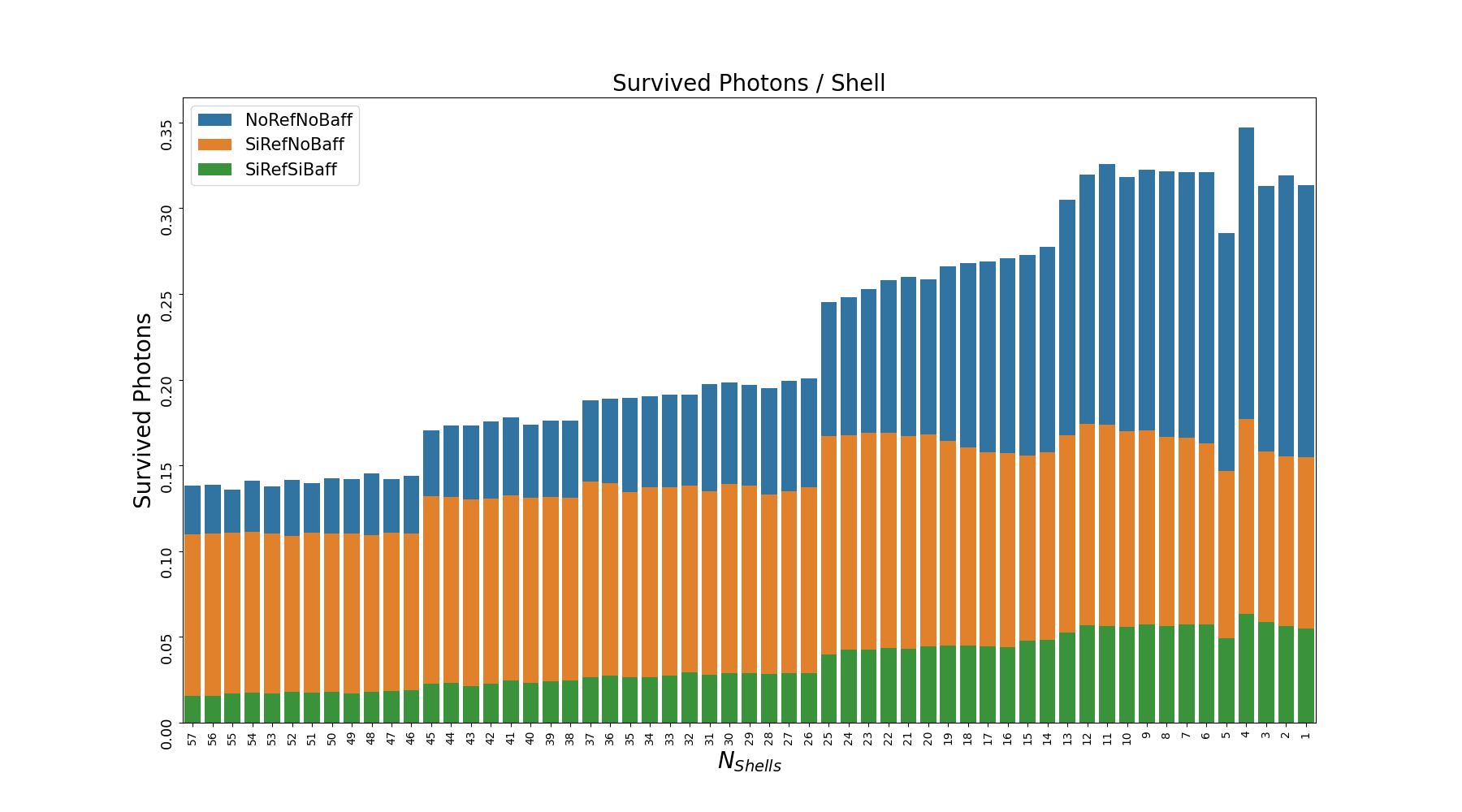}
    \caption{Representation of the different impacts of the configurations considered on stray light rejection. First, the telescope's geometrical configuration, defined by the spacing between the shells and the region of reflections, obscures 65\% to 85\% of the photons depending on the shell. The addition of reflectivity and the baffle increases the rejection power of the system, consistent with instrumental testing.}
    \label{fig:photonsurvivals}
\end{figure}

\subsection{Reflectivity and baffles}
When photons interact with the \textit{XMM-Newton} mirrors, their probability of being reflected depends on their energy and the incidence angle. The mirrors are coated with a thin layer of gold \citep{dechambure1997producing}, which determines this reflectivity, denoted as $\eta(\theta, E)$. Reflectivity plays a crucial role in determining the fraction of off-axis photons contributing to the stray light pattern, as photons from large off-axis angles are reflected with lower efficiency.

We evaluated the impact of reflectivity and baffle on the stray light pattern by simulating the following configurations:
\begin{itemize}
    \item No reflectivity, no baffle
    \item Reflectivity, no baffle
    \item Reflectivity and baffle
\end{itemize}

Our results show that neither reflectivity nor the baffle alter the geometrical pattern of the arcs, but they significantly reduce the flux (see Fig. \ref{fig:fluxcomp}). The individual contribution of reflectivity and baffle can be observed in Fig. \ref{fig:photonsurvivals}, showing that the mirror module geometry alone rejects 65-85\% of stray light photons. Adding reflectivity and baffle increases this rejection to over 95\%.

Our baffle implementation is an approximation, inspired by the collimator model described in \cite{mori2022collimators}, with heights uniformly distributed in the [25, 160] mm interval \citep{de1999x}. The addition of the collimator reduces the total stray light flux by a factor of 4.83, corresponding to a 74\% reduction compared to the reflectivity-only case. This suppression is consistent with the expected performance of the \textit{XMM-Newton} baffle system \citep{de1999xmm, de1999x}.

\newpage

\section{Mathematical derivations} 
\label{appendix:AppendixC}

\subsection{Deriving misalignment equations}
In this section we derive Eqs. \eqref{eqmis1} and \eqref{eqmis2} that describe the stray-light pattern of a mirror shell, whose misalignment is specified by the two angles $(\gamma,\xi)$ as shown in Fig. \ref{fig:tilt1}. We begin with a simplified case study involving a point source whose position is defined by the angular coordinates $(\theta, \iota)$. We initially assume the source to lie at $\iota = 0$ in the $(x, y, z)$ reference frame, while maintaining a generic $\theta$ that satisfies the conditions necessary for the generation of stray light arcs. Within this frame, the direction of rays coming from the source is expressed as:
\begin{align*}
    \vec{k_0} &= \begin{bmatrix}
           - \sin\theta \\
           0 \\
           -\cos\theta
         \end{bmatrix}
\end{align*}
We now operate a rotation to bring the optical axis in the $(x,z)$ plane

\begin{align*}
    \vec{k'_0} &= \begin{bmatrix}
            \cos\xi & \sin\xi & 0 \\
           -\sin\xi & \cos\xi & 0\\
           0 & 0 & 1
         \end{bmatrix}
         \begin{bmatrix}
           - \sin\theta \\
           0 \\
           -\cos\theta
         \end{bmatrix} = 
         \begin{bmatrix}
           - \sin\theta \cos\xi \\
            \sin\theta \sin\xi \\
           -\cos\theta
         \end{bmatrix}.
\end{align*}
A rotation of $-\gamma$ around the $y-$axis orients the shell along the $z-$axis
\begin{align*}
    \vec{k''_0} &= \begin{bmatrix}
            \cos\gamma & 0 & -\sin\gamma \\
           0 & 1 & 0\\
           \sin\gamma & 0 & \cos\gamma
         \end{bmatrix}
         \begin{bmatrix}
           - \sin\theta \cos\xi \\
           \sin\theta \sin\xi \\
           -\cos\theta
         \end{bmatrix} = \\
          &= \begin{bmatrix}
           - \sin\theta \cos\xi \cos\gamma + \sin\gamma \cos\theta \\
            \sin\theta \sin\xi \\
           - \sin\theta \cos\xi \sin\gamma - \cos\gamma\cos\theta
         \end{bmatrix}.
\end{align*}
This equation can be simplified in the small-angle approximation for $\gamma$ and $\theta$ to be small

\begin{align*}
    \vec{k''_0} &= \begin{bmatrix}
           - \theta \cos\xi + \gamma \\
           \theta \sin\xi \\
           -1
         \end{bmatrix}.
\end{align*}
In this reference frame the surface of the mirror is a cone with the axis parallel to the $z-axis$ but rotated by $\xi$

\begin{align*}
    \vec{r''_1} &= \begin{bmatrix}
            r_1 \cos(\phi - \xi) \\
            r_1 \sin(\phi - \xi) \\
            (r_1 - R_0)/\tan \beta_0
         \end{bmatrix}
\end{align*}
with $r < R_0$. In conical approximation, the normal, unit vector to the hyperboloid is 
\begin{align*}
    \vec{n''} &= \begin{bmatrix}
            - \cos \beta_0\cos(\phi - \xi) \\
            - \cos \beta_0\sin(\phi - \xi) \\
            \sin \beta_0 
         \end{bmatrix} \simeq 
         \begin{bmatrix}
             -\cos(\phi - \xi) \\
             -\sin(\phi - \xi) \\
             \beta_0             
         \end{bmatrix}
\end{align*}
and the direction of the reflected ray becomes \citep{spiga2009analytical}
\begin{align*}
    \vec{k''_1} &= \vec{k''_0} - 2(\vec{k''_0} \cdot \vec{n''}) \vec{n''} = \\
    &= \begin{bmatrix}
             \theta\cos(2\phi - \xi) - \gamma\cos(2\phi - 2\xi) -2\beta_0\cos(\phi-\xi) \\
             \theta\sin(2\phi - \xi) - \gamma\sin(2\phi - 2\xi) -2\beta_0\sin(\phi-\xi)\\
             -1          
        \end{bmatrix}.
\end{align*}
We can now return to the reference from of the misaligned shell, by operating another rotation by a $\gamma$ angle

\begin{multline*}
    \vec{k'_1} = \begin{bmatrix}
        \cos\gamma & 0 & \sin\gamma \\
           0 & 1 & 0\\
           -\sin\gamma & 0 & \cos\gamma
    \end{bmatrix} \\
    \begin{bmatrix}
             \theta\cos(2\phi - \xi) - \gamma\cos(2\phi - 2\xi) -2\beta_0\cos(\phi-\xi) \\
             \theta\sin(2\phi - \xi) - \gamma\sin(2\phi - 2\xi) -2\beta_0\sin(\phi-\xi)\\
             -1          
    \end{bmatrix} = \\
    \begin{bmatrix}
        \theta \cos(2\phi - \xi) - \gamma[\cos (2\phi - 2\xi) + 1] - 2\beta_0 \cos(\phi - \xi) \\
        \theta \sin(2\phi - \xi) - \gamma\sin (2\phi - 2\xi) - 2\beta_0 \sin(\phi - \xi) \\
        -1
    \end{bmatrix}
\end{multline*}

and another rotation to re-align the x-axis to its original orientation:

\begin{multline*}
    \vec{k_1} = \begin{bmatrix}
            \cos\xi & -\sin\xi & 0 \\
           \sin\xi & \cos\xi & 0\\
           0 & 0 & 1
         \end{bmatrix} \\
         \begin{bmatrix}
        \theta \cos(2\phi - \xi) - \gamma[\cos (2\phi - 2\xi) + 1] - 2\beta_0 \cos(\phi - \xi) \\
        \theta \sin(2\phi - \xi) - \gamma\sin (2\phi - 2\xi) - 2\beta_0 \sin(\phi - \xi) \\
        -1
    \end{bmatrix} = \\
    \begin{bmatrix}
        \theta \cos2\phi - \gamma[\cos(2\phi - \xi) + \cos\xi] - 2\beta_0\cos\phi \\
        \theta \sin\phi - \gamma[\sin(2\phi - \xi) + \sin\xi] - 2\beta_0\sin\phi \\ 
        -1
    \end{bmatrix}.
\end{multline*}
The reflected rays intersect the focal plane at the coordinates $z=
-f$, i.e. at the point:
\begin{equation*}
    \vec{r} = \vec{r_1} + f\vec{k_1}
\end{equation*}
obtaining the vector
\begin{align*}
    \begin{bmatrix}
        x \\
        y \\
        z 
    \end{bmatrix} &=
    \begin{bmatrix}
        \theta f \cos2\phi - \gamma f[\cos (2\phi - \xi) + \cos\xi] + (R_0 - 2\beta_0 f)\cos\phi \\
        \theta f \sin2\phi - \gamma f[\sin (2\phi - \xi) + \sin\xi] + (R_0 - 2\beta_0 f)\sin\phi \\
        -f
    \end{bmatrix}
\end{align*}
and so we obtain the parametric equations
\begin{gather*}
    x(\phi) = (R_0 - 2\beta_0 f)\cos\phi + \theta f \cos2\phi - \gamma f[\cos (2\phi - \xi) + \cos\xi]  \\
    y(\phi) = (R_0 - 2\beta_0 f)\sin\phi + \theta f \sin2\phi - \gamma f[\sin (2\phi - \xi) + \sin\xi]
\end{gather*}.

These equations can be generalized with $\xi \rightarrow \xi - \iota$ including the case where $\iota \neq 0$, operating another rotation of the reference frame by an angle $\iota$ around the z-axis, for aligning the source to its correct direction.

\subsection{Deriving shift equations}
We derive Eqs. \eqref{shift1} and \eqref{shift2} by assuming a random shift of the optical axis with amplitude $\delta$ in the $\epsilon$ orientation on the aperture plane. The coordinates of the source with respect to the new optical axis can be expressed with the following equations
\begin{gather*}
    x' = x + dx \\
    y' = y + dy
\end{gather*}
where 
\begin{gather*}
    dx = \left(\frac{\delta}{f}\right)\cos\epsilon \\ 
    dy = \left(\frac{\delta}{f}\right)\sin\epsilon.
\end{gather*}
We write $(x,y)$ with respect to the polar coordinates
\begin{gather*}
    x = \sin\theta\cos\iota \\
    y = \sin\theta\sin\iota
\end{gather*}
and substitute in the previous equations. 
\begin{gather*}
    x' = \sin\theta\cos\iota + \left(\frac{\delta}{f}\right)\cos\epsilon \\
    y' = \sin\theta\sin\iota + \left(\frac{\delta}{f}\right)\sin\epsilon.
\end{gather*}
Equations \eqref{shift1} and \eqref{shift2} can be obtained by transforming these coordinates in polar coordinates
\begin{gather*}
    \theta' = \sqrt{x'^2 + y'^2} \\
    \iota' = \arctan\left(\frac{y'}{x'}\right)
\end{gather*}

\end{appendix}

\end{document}